\documentclass[conference]{IEEEtran}

\IEEEoverridecommandlockouts

\makeatletter
\def\ps@headings{%
\def\@oddhead{\mbox{}\scriptsize\rightmark \hfil \thepage}%
\def\@evenhead{\scriptsize\thepage \hfil \leftmark\mbox{}}%
\def\@oddfoot{}%
\def\@evenfoot{}}
\makeatother
\usepackage{amsmath}
\usepackage{graphicx}
\usepackage{amssymb}
\usepackage{bm}
\usepackage{ifthen}
\usepackage{paralist}
\usepackage{cite}
\usepackage{url}
\usepackage{subfigure}
\usepackage{times}
\usepackage{color}

\newboolean{singlever}
\setboolean{singlever}{false}

\ifthenelse{\boolean{singlever}}{
  \renewcommand{\baselinestretch}{1.45}
  \addtolength{\abovecaptionskip}{-0.27cm}
  \addtolength{\belowcaptionskip}{-0.27cm}
}

\usepackage{times}

\newcommand{\set}[1]{\ensuremath{\mathcal #1}}

\newcommand{\separator}{
  \begin{center}
    \rule{\columnwidth}{0.3mm}
  \end{center}
}

\newtheorem{theorem}{Theorem}[section]

\newtheorem{proposition}{Proposition}[section]

\newtheorem{definition}{Definition}[section]

\newtheorem{remark}{Remark}[section]

\newcommand{\expect}[1]{\mathbb{E}[ #1 ]}

\newcommand{\beq}{\begin{eqnarray*}}
\newcommand{\eeq}{\end{eqnarray*}}
\newcommand{\beqn}{\begin{eqnarray}}
\newcommand{\eeqn}{\end{eqnarray}}
\newcommand{\bemn}{\begin{multiline}}
\newcommand{\eemn}{\end{multiline}}

\begin{document}
\title{Economics of WiFi Offloading: \\Trading Delay for Cellular Capacity
}

\author{ Joohyun~Lee$^\dag$, 
  Yung Yi$^\dag$, 
  Song Chong$^\dag$, and Youngmi Jin$^\dag$\thanks{$^\dag$The authors are with Electrical
    Engineering, KAIST (Korea Advanced Institute of Science and
    Technology, e-mails: jhlee@netsys.kaist.ac.kr, \{yiyung,
    songchong\}@kaist.edu,
    youngmi\_jin@kaist.ac.kr} 
}


\maketitle

\begin{abstract}
  Cellular networks are facing severe traffic overloads due to the
  proliferation of smart handheld devices and traffic-hungry
  applications. A cost-effective and practical solution is to offload
  cellular data through WiFi. Recent theoretical and experimental
  studies show that a scheme, referred to as delayed WiFi offloading,
  can significantly save the cellular capacity by delaying users' data
  and exploiting mobility and thus increasing chance of meeting WiFi APs
  (Access Points). Despite a huge potential of WiFi offloading in
  alleviating mobile data explosion, its success largely depends on the
  economic incentives provided to users and operators to deploy and use
  delayed offloading. In this paper, we study how much economic benefits
  can be generated due to delayed WiFi offloading, by modeling a market
  based on a two-stage sequential game between a monopoly provider and
  users.  We also provide extensive numerical results computed using a
  set of parameters from the real traces and Cisco's projection of
  traffic statistics in year 2015. In both analytical and numerical results,
  we model a variety of practical scenarios and control knobs in terms
  of traffic demand and willingness to pay of users, 
  spatio-temporal dependence of pricing and traffic, 
  and diverse pricing and delay tolerance.
  We demonstrate that
  delayed WiFi offloading has considerable economic benefits,
  where the increase ranges from
  21\% to 152\% in the provider's revenue, and from 73\% to 319\% in the
  users' surplus, compared to on-the-spot WiFi offloading.

\end{abstract}




\label{sec:introduction}
\section{Introduction}


Mobile data traffic is growing enormously, as smart phones/pads equipped
with high computing powers and diverse applications are becoming popular.
Cisco reported that global mobile data traffic grew 2.3-fold in 2011,
more than doubling for the four consecutive years, which supports its
previous annual forecast since 2008~\cite{cisco12}.  It was also
forecast there that the total global mobile data traffic will increase
18-fold between 2011 and 2016, where the average smartphone is projected
to generate 1.3 GB per month in 2015~\cite{cisco12}.
To cope with such mobile data explosion, upgrading to 4G
(e.g., LTE (Long Term Evolution) or WiMax), may be an immediate
solution, but mobile applications are becoming more diverse with larger
data consumption and the number of smartphone/pad users are also
increasing rapidly. Then, users' traffic demand is expected to exceed
the capacity of 4G in the near future, and thus mobile network 
operators\footnote{We use `operator' and `provider' interchangeably throughout
  this paper.} keep seeking other alternatives to efficiently
respond to data explosion \cite{4g_not_solution,lte_not_solution}.



WiFi offloading, where users use WiFi prior to 3G/4G whenever they have
data to transmit/receive, has been proposed as a practical solution that
can be applied without much financial burden in practice. Network
operators as well as users can easily and quickly install WiFi 
access points (APs) with low costs, and in fact many operators worldwide
have already deployed and provided WiFi services in hot-spots and
residential areas.  Recent papers \cite{conext10,offloading:mobisys10} 
demonstrate that a huge portion of cellular traffic can be offloaded to
WiFi by letting users delay their delay-tolerant data
(e.g., movie, software downloads, cloud backup and sync services),
and upload/download data whenever they meet a WiFi AP within a pre-specified
delay deadline. We call this {\em delayed WiFi offloading}, and about
60-80\% of cellular traffic can be reduced when 30 mins to 1 hour
delay for human mobility \cite{conext10} and 10 mins of delay for
vehicular mobility \cite{offloading:mobisys10} are allowed.  This remarkable
offloading efficiency is due to users' mobility enabling themselves to
be under a WiFi coverage during a considerable portion of their business
time.  Example usage scenarios include: 1) Alice records video of a family
outing at a park using her cell phone and wants to archive it in her
data storage in the Internet. She does not need the video immediately
until she comes back home in several hours.
2) Bob will travel this afternoon
from New York to Los Angeles and he just realizes that
he can use some entertainment during the long flight. As he has
several hours before the trip, he schedules to download a couple
of movies on his cell phones. 



However, WiFi offloading's high potential does not always guarantee that
users and providers adopt it in practice. First, users may be reluctant
to delay their traffic without economic incentives, e.g., discounted
service fees. For example, if a user pays based on an unlimited data
plan, which is still a popular payment plan worldwide, users may have no
reason to delay traffic unless WiFi-required services, e.g., services
requiring higher bandwidths, are necessary. Also, operators may not
always welcome delayed offloading service, since the total cellular
traffic to charge may decrease, possibly leading to its revenue
reduction. Thus, it is of significant importance to formally address the
question on the economic gains of delayed WiFi offloading 
from the perspective of users, operators,
and regulators, which is the focus of this paper.



In this paper, we model a market with a monopoly operator and users based on a
two-stage sequential game, where the operator controls the price and
users are price-takers. A variety of control knobs will show different
economic impacts of delayed WiFi offloading. Our major focus is to
understand how and how much users and the provider obtain the economic
incentives by adopting delayed WiFi offloading and study the effect of
different pricing and delay-tolerance. 
The major features of our model include four different pricing schemes
(flat, volume, two-tier, and congestion) and heterogeneous users in terms
of traffic demands and willingness to pay.

Using the market model mentioned above, we first conduct analytical
studies under flat and volume pricing for the simple cases when
the traffic demand follows a certain distribution (obtained from the
measurement studies), and users are uniformly distributed among
cells. This simplification seems to be unavoidable for mathematical
tractability, yet we are able to fundamentally understand how the users
and the provider become economically beneficial.
We formally prove that delayed WiFi offloading indeed generates the
economic incentives for the users and the provider. To obtain more
practical messages and quantify the gain of delayed WiFi offloading, we
use two traces, each of which tells us the information on cellular data
usage and WiFi connectivity. We extract the parameters needed by our
model from those traces, and obtain numerical results, from which, we
draw the following key messages:

\smallskip
\begin{compactenum}[(a)]
\item WiFi offloading is economically beneficial for both the provider
  and users, where depending on the pricing schemes and delay tolerance, 
  the increase ranges from 21\% to 152\% in the provider's revenue
  and from 73\% to 319\% in the users' surplus.
\item Revenue in volume pricing exceeds that in flat pricing. 
  However, the revenue {\em increasing rate} of delayed offloading in flat pricing
  is higher than that in volume pricing.
\item Pricing with higher granularity such as two-tier and congestion
  pricing increases the revenue, compared to flat and volume pricing,
  but the gains 
  become smaller as offloading efficiency increases (i.e.,
  as users delay more traffic).

\item The revenue gain from on-the-spot to delayed offloading is similar
  to that generated by the network upgrade from 3G to 4G.







\end{compactenum}

\label{sec:related_work}
\section{Related Work}


There have been several works~\cite{conext10,offloading:mobisys10,serviceprovisioning}
on delayed WiFi offloading.
Lee et al.~\cite{conext10} proposed a delayed offloading framework,
where users specify a deadline for each application or data,
and each delay tolerant data is served in a shortest remaining time first (SRTF) manner
through WiFi networks.
If the delay deadline of the data expires, then the data is transmitted through 3G networks.
On real human mobility traces, it is shown that 80\% of cellular traffic
can be offloaded to WiFi networks when 1 hour delay is allowed. 
Balasubramanian et al.~\cite{offloading:mobisys10} 
proposed an offloading framework for vehicular networks,
which supports fast switching between 3G and WiFi,
and avoids bursty WiFi losses. 
They demonstrated that more than 50\% of cellular traffic can be 
reduced for a delay tolerance of 100 seconds on vehicular mobility. 
Hultell et al.~\cite{serviceprovisioning} addressed the user experienced
performance of delayed transmission and proposed context-aware
caching/prefetching to provide users immediate services, e.g. web
browsing, news, and streaming.
They found that more than 80\% of news can be pre-fetched within 700 
seconds, with only 50 WiFi APs per km$^2$ (1.2\% spatial coverage) 
on a random mobility model. Therefore, WiFi networks are proven to
offload large fraction of cellular data for various mobility conditions 
and low AP density, whenever data can tolerate some amount of delay. 

Some recent works~\cite{tube_sangtae,win_coupon} devised incentive frameworks 
for users to delay their data traffic.
Ha et al.~\cite{tube_sangtae} proposed a time-dependent pricing scheme 
for mobile data, which incentivizes users to delay their traffic 
from the higher- to lower-price time zone.
They conducted surveys which revealed that users are indeed willing to wait
5 minutes (for YouTube videos) to 48 hours (for software updates). 
They addressed that the time-dependent pricing flattens temporal fluctuation of
traffic usage and benefits wireless operators.
In~\cite{win_coupon}, Zhuo et al. proposed an incentive framework for 
downlink mobile traffic offloading based on an auction mechanism,
where users send bids, which include the delay it can tolerate and 
the discount the user wants to obtain for that delay, 
and the provider buys the delay tolerance from the users.
However, previous studies did not provide how much economic gain 
the provider and users can obtain. 
In this paper, we quantify the economic gain of delayed offloading
based on real-world traces. 






\begin{figure}[t!]
  \center
  \includegraphics[width=1\columnwidth]{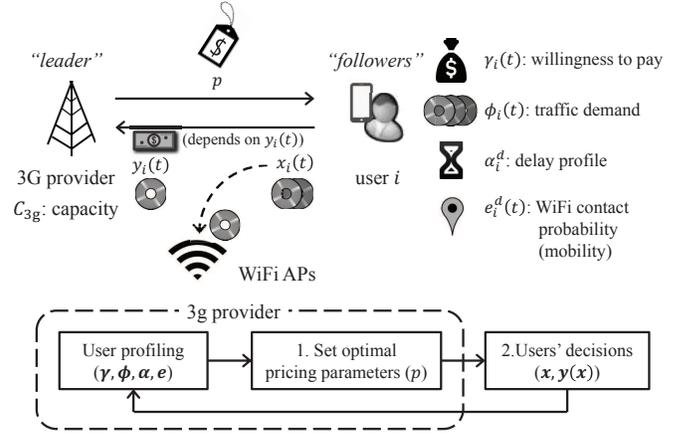}     
  \caption{
    An illustration of the system model. 
    $x_i(t)$ and $y_i(t)$ are 3G+WiFi and 3G traffic volumes of user $i$ at time $t.$
    Note that $x_i(t) \leq \phi_i(t).$ 
    \label{fig:illustration}}
\end{figure}

\section{Model}
\label{sec:system_model}

We illustrate the system model in Fig.~\ref{fig:illustration}.
We model users with four attributes, 
{\em (i)} how much money they can pay ({\em willingness to pay}, $\bm{\gamma}$),
{\em (ii)} how much data they want to use ({\em traffic demand}, $\bm{\phi}$),
{\em (iii)} how long their data can tolerate ({\em delay profile}, $\bm{\alpha}$),
and {\em (iv)} how they move ({\em WiFi contact probability}, $\bm{e}$).
We index users by $i$, time slots by $t$, and deadlines by $d.$
Assuming that the monopoly provider knows users' attributes and strategies a priori,
we model a market model based on a two-stage sequential game 
(e.g. Stackelberg game).
At the first stage, the provider decides on the pricing parameters ($p$)
(for a fixed pricing scheme, which we will describe later)
as a {\em leader}, and at the second stage, each user is a price-taker
as a {\em follower} and chooses the 3G+WiFi traffic volume $\bm{x}.$
Our analysis and numerical results are carried out based on the equilibrium of this game.
We describe the detailed network, traffic, and market models in the following subsections.
We summarize major notations in Table~\ref{tbl:notations}.

\subsection{Network and Traffic Model}

\subsubsection{Network model}
We consider a network consisting of cellular base stations
(BSs) and WiFi APs, where $N$ users are served by the
cellular provider.\footnote{Throughout this paper, we use the words `BS'
  and `AP' to refer to a cellular BS and a WiFi AP, respectively.}
Users are always guaranteed to be under the coverage of a cellular BS,
but not necessarily of a WiFi AP.
We consider a one-day time scale whose average analysis
over the unit billing cycle, e.g., one month, is presented.  
A day is divided into time slots $t \in \{1, 2, \ldots, T\},$ where $T$ is the last
index of one day, depending on the duration of a time slot.\footnote{We
  also use $N$ and $T$ to refer to a set of all users and time slots to
  abuse the notation.}  Let $C_{\text{3g}}$ be the capacity (in volume per
slot) provided by a BS.  During each day, users move among BSs
as well as APs.  Let $e_{i}^d(t)$ be the probability that user $i$ meets
any WiFi AP within deadline $d$ at time slot $t.$ For instance,
$e_{i}^{\text{1hour}}(\text{13:00}) = 0.7$ means that user $i$ meets a
WiFi AP from 1 p.m. to 2 p.m. with probability 0.7.  The value of
$e_{i}^d(t)$ can be obtained by analyzing user $i$'s mobility trace
during, say, a month.
We assume that only 3G traffic is charged. Each user has its own set of
accessible, free WiFi APs, e.g., ones in home, office, or hotspots,
deployed by users, users' companies, providers, or governments.
We ignore the cost from offloaded data
since the cost of accessing the Internet via a WiFi AP connected to
a wired network is considerably lower than that for accessing the cellular network~\cite{cellular_cost}. 


\begin{table}[t!]
  \caption{Summary of Major Notation}
  \vspace{-0.2cm}
  \begin{centering}
    \small
    \begin{tabular}{|c|l|}
      \hline
      Variable & Definition\\
      \hline
      \hline
      $N$ and $C_{\text{3g}}$ & Number of users and Capacity of a BS cell\\
      \hline
      $\theta$& Price sensitivity \\
      \hline
      $e_{i}^d(t)$ & The WiFi contact probability of user $i$ \\
       & at slot $t$ within deadline $d$\\      
      \hline
      $\alpha_{i}^d $ & Portion of traffic of user $i$ with deadline $d$\\
      \hline
      \hline
      $x_i (t)$& 3G+WiFi traffic volume of user $i$ at slot $t$\\
      \hline      
      $y_i (t)$& 3G traffic volume of user $i$ at slot $t$\\
      \hline
      $\phi_i(t)$ & Traffic demand of user $i$ at slot $t$ \\
      \hline
      $w_i(t)$ & Temporal preference (weight) of user $i$ at time $t$\\
      \hline
      $\gamma_i (t)$ & Willingness to pay of user $i$ for traffic at time $t$\\
      \hline
    \end{tabular}\label{tbl:notations}
    \par\end{centering}
\end{table}

\subsubsection{Traffic model}
We assume that user $i$ has the average daily traffic demand ${\Phi}_i,$
3G+WiFi traffic vector $\bm{x}_i = (x_i(t): t\in T),$ and
3G traffic vector $\bm{y}_i(\bm{x}_i) = (y_i(t): t\in T),$
where $x_i(t)$ is the traffic volume of user $i$ generated at slot $t$,
transferred through either 3G or WiFi,
and $y_i(t)$ is the traffic volume transferred through only 3G. 
The daily traffic demand ${\Phi}_i$ is
temporally split into $\bm{\phi}_i=(\phi_i(t): t \in T),$ where
$\phi_i(t)$ is traffic demand at slot $t,$ and ${\Phi}_i = \sum_{t \in
  T} \phi_i(t).$
The traffic volume of user $i$ at slot $t$ is constrained
by the traffic demand, i.e., $x_i(t) \leq \phi_i(t).$
We denote $w_i(t) = \frac{\phi_i(t)}{{\Phi}_i}$
as temporal preference (weight) of user $i$.  
For example, for a user
$i$'s traffic demand 1 GB, where users want to send 700 MB at daytime an
300 MB at nighttime (i.e., just two time slots), $\phi_i(\text{day}) =
700$ and $\phi_i(\text{night}) = 300,$ and $w_i(\text{day})=0.7$ and
$w_i(\text{night})=0.3.$
Users may not be able to deliver all traffic demand, 
and the actual transmitted volume depends on the price and user utility,
which we will describe later.
The traffic volume for 3G, $\bm{y}_i$
which is actually charged, relies on $\bm{x}_i$ as well as each user
$i$'s mobility and delay profile, which we define in the following paragraph.



We introduce a notion of {\em delay profile} to model per-user
delay-tolerance of traffic. 
The delay profile is denoted by $\bm{\alpha} =
(\alpha_{i}^d: i \in N, d \in \{0, 1, ... , D\}) $ such that
$\sum_{d=0}^{D} \alpha_{i}^d = 1,$ where $\alpha_{i}^d$ is the portion
of user $i$'s traffic demand that allows deadline $d$, and $D$ is the maximum
allowable deadline across all traffic. 
For example, for a user $i$'s
traffic demand 1 GB, if the user has 300 MB, 700 MB, allowing
10 mins and 1 hour, resp., we have $\alpha_{i}^{\text{10m}} = 0.3,$ 
$\alpha_{i}^{\text{1h}} = 0.7.$
For example, for a user $i$'s
traffic demand 1 GB, if the user has 100 MB, 200 MB, 700 MB, allowing
real-time, 10 mins, and 1 hour, respectively, we have $\alpha_{i}^0 =
0.1,$ $\alpha_{i}^{\text{10m}} = 0.2,$ $\alpha_{i}^{\text{1h}} = 0.7.$
For a given per-user delay profile, each user uses only WiFi connections
to deliver some data until the allowable deadline expires, after which
the remaining data is immediately transferred through 3G.  In
particular, when no delay is allowed ($\alpha_i^0 = 1$), we
call this regime {\em on-the-spot} offloading,
where a user only uses spontaneous connectivity of WiFi.
Most current smartphones support this by default.






\subsection{Market Model}

We start by explaining the economic metrics of the users and the
provider. We assume that the provider and users are rational and try to maximize revenue
and net-utility.

\subsubsection{Users and Provider}

We model heterogeneous willingness to pay among users over time slots,
which we denote by $\gamma_i(t) \geq 0 $ for user $i$ at time $t.$
For an average user $i,$ $\gamma_i(t)$ tends to be higher when $t$ is in daytime.
We first define user $i$'s utility at time slot $t$ by $\gamma_i(t)
x_i(t)^{\theta},$
where the constant $\theta \in (0,1)$ is price-sensitivity. The utility
function $\gamma_i(t) x_i(t)^{\theta}$ is called an {\em iso-elastic}
function\footnote{A function $u(x)$ is said to be \emph{iso-elastic} if
  for all $k > 0$, $u(kx) = f(k)u(x) + g(k)$ for some functions $f(k),
  g(k)>0$.} with the property of an increasing function of     
traffic volume $x_i(t)$ for all $i$ and $t$, but of a decreasing
marginal payoff.
Then, user $i$'s (aggregate) net-utility $U_i(\bm{x}_i)$ during a day is:
\begin{eqnarray*}
  U_i(\bm{x}_i)  =  \sum_{t \in T} \gamma_i(t) x_i(t)^{\theta}- m\big(p, \bm{y}_i
  (\bm{x}_i)\big),
\end{eqnarray*}
where $m(p, \bm{y}_i(\bm{x}_i))$ is the daily payment charged by the
provider whose price is $p$.  We abuse the notation and use $p$ to refer
to the pricing parameters of a given pricing scheme, and the function
form of $m$ differs depending on a pricing scheme (see
Section~\ref{sec:pricing_model}).  Recall the notation $\bm{y}_i
(\bm{x}_i)$ represents the dependency of the 3G traffic on 3G+WiFi
traffic.

Given the traffic demand $\bm{\phi}_i,$ mobility pattern, willingness
to pay $\bm{\gamma}_{i}=(\gamma_i(t): t \in T)$, delay profile $\alpha_i^d$,
and a pricing scheme (and its parameters), each user $i$ chooses
$\bm{x}^\star_i$ to maximize his/her net-utility, 
\begin{eqnarray}
  \label{eq:user}
  {\bf User:} && \displaystyle  \max_{{x}_i(t) \leq \phi_i(t), \ \forall t
  \in T} U_i(\bm{x}_i), 
\end{eqnarray}
where 
each user $i$ subscribes to 3G service only if the net-utility is
positive, i.e., $ U_i(\bm{x}_i) > 0.$


Under a given pricing scheme, the provider decides on the price (more
precisely, the parameters of the pricing scheme) to maximize its
expected revenue, $R(p)$:
\begin{eqnarray}
  \label{eq:provider}
  {\bf Provider:} && \displaystyle \max_{p \in \set{P}} R(p), 
\end{eqnarray}
where $\set{P}$ is the set of
all feasible prices such that
{\em (i)} the revenue is positive ({\em provider rationality}), and
{\em (ii)} the expected 3G traffic volume
at each time and at each BS cell is smaller than 3G capacity
$C_{\text{3g}}$ ({\em capacity constraint}).

The expected revenue $R(p)$ is the total {\em income} minus {\em cost}, 
\begin{eqnarray}
  \label{eq:revenue}
  R(p) = \sum_{i \in N} m\big(p, \bm{y}_i(\bm{x}_i)\big) -
  \sum_{i \in N} c\big(\bm{y}_i(\bm{x}_i)\big),
\end{eqnarray}
where $c(\bm{y}_i)$ is the network cost to handle the 3G traffic, which
we model by a linearly increasing function, $c(\bm{y}_i) = \eta \sum_{t
  \in T} y_i(t),$ where $\eta$ is the cost of the unit volume of
the 3G traffic.  The cost term captures the money for operation and
maintenance including electric power costs as well as customer complaints
due to congestion.  The linearly increasing network cost is commonly
used in the analysis of cellular cost \cite{linear_network_cost}.  
We ignore the backhaul cost of both 3G and WiFi networks.
User surplus $S$ is the summation of users' net-utility and social
welfare $W$ is the summation of user surplus and provider revenue, or,
\ifthenelse{\boolean{singlever}}{
\begin{eqnarray*}
  S = \sum_{i \in N} U_i (\bm{x}_i), \qquad
  W = \sum_{i \in N, t \in T} \gamma_i(t) x_i(t)^{\theta}- \sum_{i \in N} c\big(\bm{y}_i(\bm{x}_i)\big).
\end{eqnarray*}
}
{
\begin{eqnarray*}
  S &=& \sum_{i \in N} U_i (\bm{x}_i), \cr
  W &=& \sum_{i \in N, t \in T} \gamma_i(t) x_i(t)^{\theta}  - \sum_{i \in N} c\big(\bm{y}_i(\bm{x}_i)\big).
\end{eqnarray*}
}

\subsubsection{Pricing}
\label{sec:pricing_model}
For a given pricing scheme, the 3G provider fixes a {\em price
  parameter} which is announced to the users. We consider four pricing
schemes - {\em flat, two-tier, volume}, and {\em congestion} - that are
popularly studied in literature. Each pricing scheme has tunable
parameters controlled by the provider: $\{p_f\},~\{p_t^1,
p_t^2,y_{\text{max}}^1\},$ $\{p_v\},$ and $\{p_v(t,s)\},$ 
which we elaborate shortly. For a given pricing scheme
and its price parameters $p$, a user with 3G traffic volume $\bm{y}_i$ pays
$m\big(p,\bm{y}_i\big)$ to the provider.
Note that if a user does not subscribe or generate any traffic,
the payment is zero, i.e., $m\big(p,\bm{0}\big) = 0.$

\smallskip
\noindent {\bf \em Flat.} The provider offers unlimited service for
users who pay a subscription fee $p_f$.


\noindent {\bf \em Two-tier.}
Multiple price points are provided for several usage options.
For example, AT\&T has a pricing plan that offers up to
300 MB, 3 GB, and 5 GB for \$20, \$30, and \$50 per month, respectively.
In this paper, we consider two price points, where the provider offers
maximum daily traffic volume $y^1_{\max}$ for fixed fee $p_t^1$ and
unlimited service for fixed fee $p_t^2$, or,
\begin{eqnarray*}
  m\big(p,\bm{y}_i\big) &=&
  \begin{cases}
    p_t^1, & \mbox{if } \quad 0 < \sum_{t \in T} y_i (t) \leq y^1_{\max}. \\
    p_t^2, & \mbox{if } \quad \sum_{t \in T} y_i (t) > y^1_{\max}.
  \end{cases}
\end{eqnarray*}


\noindent   {\bf \em Volume.}
A user is charged to pay $p_v$ for the unit 3G traffic volume, or,
\begin{eqnarray*}
  m\big(p,\bm{y}_i\big) &=& p_v \cdot \sum_{t \in T} y_i (t).
\end{eqnarray*}

\noindent {\bf \em Congestion.}
We consider a volume-based congestion pricing, or simply congestion
pricing in this paper, where the price for a unit file size 
varies with time and location, or,
\begin{eqnarray*}
  m\big(p,\bm{y}_i\big) = \sum_{t \in T} {p_v (t,s_i(t)) \cdot y_i (t)},
\end{eqnarray*}
where 
$p_v (t,s)$ is the unit price at time slot $t$ and cell $s,$ and
$s_i(t)$ is the cell id with which user $i$ is associated at $t.$

\smallskip
We note that 
in flat and two-tier 
pricing, users do not subscribe to the
service when the net-utilities are not positive, 
which is the major factor determining the provider's revenue, 
whereas in volume and congestion 
pricing, every user subscribes and just controls its
traffic volume. 
Two-tier and congestion pricing schemes are the
extensions of flat (in terms of price granularity) and volume (in terms
of space and time), resp.  Tiered pricing in mobile data
services is popularly used recently~\cite{cisco12} as the provider can
set up multiple pricing points, while maintaining 
simplicity in the pricing structure.  
Congestion pricing has been considered as a way of revenue increase 
in networking services 
(see e.g., \cite{tube_sangtae,tsitsiklis00,pricing_congestible}). 
In fact, in the usage of cellular networks,
it has been reported that 
spatial and temporal variation of mobile data
traffic are shown to be remarkable \cite{cellular_traffic}, implying high potential
in the increase of the provider's revenue and better resource utilization. 
Despite the high billing complexity, it is interesting to
see its quantified 
impact in WiFi offloading.

\section{Analysis of WiFi Offloading Market}
\label{sec:analysis}

In Sections~\ref{sec:analysis_flat} and \ref{sec:analysis_volume}, we
provide analytical studies of the economic gain of WiFi offloading. 
Due to complex interplays among pricing parameters, and 
more importantly users' heterogeneity, our analysis is made under several
assumptions. 
This simplification seems unavoidable for mathematical tractability,
yet we are able to understand how the users and the provider become economically 
beneficial.
Note that in Section~\ref{sec:numerical_results},
we quantify economic gains of WiFi offloading
in more practical settings (heterogeneous cells and willingness to pay of users) 
as well as complex pricing schemes ({\em two-tier} and {\em congestion}).

\subsection{Assumptions and Definitions}

\begin{compactenum}[\bf {A}1.]

\item {\em Homogeneous cells.}
  User associations are uniformly distributed among
  cells so that it suffices to consider only a single BS cell, where
  the number of users is $\hat{N} = N/(\text{\# of cells})$.
  The distribution of users' traffic demand in each cell is identical.

\item {\em Traffic demand distribution.}  In each cell, the daily
  traffic demand $\Phi_i$ follows a random variable $\Phi$ which follows
  an upper-truncated power-law distribution, given by: $f_{{\Phi}} (x) =
  {x^{-\sigma}}/{Z},$ for $0 \leq x \leq {\Phi}_{\max}$, where $\sigma$
  is the exponent, ${\Phi}_{\max}$ is the maximum value of ${\Phi},$ and
  $Z = \frac{{\Phi}_{\max}^{1-\sigma}}{1-\sigma}$ with $0 < \sigma < 1.$

\item {\em Willingness to pay and temporal preference.}  Users are
  homogeneous in willingness to pay and temporal preference, 
  i.e., $\gamma_i(t) = \gamma(t), w_i(t) = w(t),~ \forall i \in N, ~ t \in T.$
  In regard to willingness to pay, we let $\gamma (t) = w(t)^
  {1-\theta}.$
  However user's traffic demand is heterogeneous as in {\bf A2}.

\item {\em Pricing.} We consider only flat and volume pricing. Thus,
  throughout this section, the pricing parameter $p$ refers to the
  flat fee $p_f$ and the unit price $p_v$ in each pricing, resp.
\end{compactenum}
\smallskip

In {\bf A2}, we comment that in recent measurement studies
\cite{cellular_traffic, cellular_traffic:sigmetrics11}, the traffic
volume distribution of cellular devices is shown to follow an
upper-truncated power-law distribution.
Especially, in~\cite{cellular_traffic},
the adopted pricing policy was flat pricing,
so that the measured traffic usage was not affected by pricing. 
Thus, we apply an upper-truncated power-law distribution to daily
traffic demand ${\Phi}.$ 
In {\bf A3}, willingness to pay $\gamma(t)$ at
time $t$ is set, such that 
at each time slot $t$ 
{(i)} one has larger
willingness to pay for larger traffic demand and 
{(ii)} utility generated by traffic demand 
($\gamma(t) \phi(t)^{\theta}$) is proportional to the traffic demand ($\phi(t) = w(t) \Phi$).


\begin{remark}
  User heterogeneity only comes from traffic demand $\Phi$ from our
  assumptions. For notational simplicity, we omit user subscript $i$ and
  use subscript ${\Phi}$ to represent the user variables with
  traffic demand ${\Phi},$ e.g., $x_{\Phi}(t), y_\Phi (t),$
  etc.
\end{remark}



\smallskip
\noindent{\bf \em Offloading indicators.} We first introduce two indicators to quantify how much 3G data is
offloaded: (i) aggregate 3G traffic ratio  $\kappa_{\text{avg}}$ and (ii) peak 3G traffic
ratio $\kappa_{\text{peak}}$.

\smallskip
\begin{definition}[offloading indicators]
\begin{eqnarray}
\label{eq:def_q_and_k}
  \kappa_{\text{avg}} \triangleq   \frac{\sum_{t \in T} Y(t)}{\sum_{t \in
      T} X(t)}, \qquad \kappa_{\text{peak}} \triangleq \frac{\max_{t \in T} Y(t)}{\sum_{t \in T} X(t)},
\end{eqnarray}
where the transmitted total traffic and 3G traffic over a cell at time $t$, $X(t)$
and $Y(t)$\footnote{When we emphasize that $Y(t)$ (resp. $X(t)$)
  depends on a given price $p$, $Y(t;p)$ will be used instead of $Y(t)$
  (resp. $X(t;p)$).} are:
\ifthenelse{\boolean{singlever}}{
\begin{eqnarray}
  X(t)  =  \hat{N} \int_0^{\Phi_{\max}} x_{\Phi}(t) \text{d} F_{\Phi},  \quad
  Y(t)  =  \hat{N} \int_0^{\Phi_{\max}}\sum_{d=0}^D b_{\Phi}^d(t-d) x_{\Phi}(t-d) \text{d} F_{\Phi}, 
  \label{eqn:Y(t)}
\end{eqnarray}
}
{
\begin{eqnarray}
  X(t) & = & \hat{N} \int_0^{\Phi_{\max}} x_{\Phi}(t) \text{d} F_{\Phi}, \nonumber\\
  Y(t) & = & \hat{N} \int_0^{\Phi_{\max}}\sum_{d=0}^D b_{\Phi}^d(t-d) x_{\Phi}(t-d) \text{d} F_{\Phi}, 
  \label{eqn:Y(t)}
\end{eqnarray}
}
and $b_{\Phi}^d(t) = \alpha_{\Phi}^d\big(1-e_{\Phi}^d(t) \big)$ 
is the portion of the traffic generated at time $t$ which is transmitted
through 3G at time $t+d.$
\end{definition}

\smallskip
\noindent
It is clear that as users delay more
traffic, the aggregate 3G traffic ratio $\kappa_{\text{avg}}$ provably decreases,
since more traffic can be offloaded through WiFi.
Also, the peak 3G ratio $\kappa_{\text{peak}}$ decreases 
as more traffic at {\em peak} time is offloaded.
In Section~\ref{sec:numerical_results}, we show that
both $\kappa_{\text{avg}}$ and $\kappa_{\text{peak}}$ decrease
as delay tolerances of users get higher. 





\smallskip
\noindent{\bf \em Opt-saturated and Opt-unsaturated.}
We define two notions, {\em opt-saturated} and {\em opt-unsaturated},
which characterize the regimes under which {\em how much traffic is
imposed on the network for the equilibrium price.} 
In general, as traffic demand gets higher compared to the 3G capacity,
the network becomes {\em opt-saturated}, and vice versa. 
The main reason for
introducing those two notions is because the analysis becomes different
depending on the volume of network traffic and the market behaves
differently, and thus, the way of increasing the revenue and the
net-utility can be differently interpreted.  For a formal definition, we
first recall that $\set{P}$ is the set of all feasible prices, defined
by provider rationality and capacity constraint, or,
\begin{eqnarray}
  \label{eq:def_feasible_set}
  \set{P} \triangleq \left\{ p \mid R(p) >  0, Y(t;p) \leq C_{\text{3g}}, ~ \forall t \in T \right\}.
\end{eqnarray}




\begin{definition}[Opt-saturated and Opt-unsaturated]
  \label{def:saturated_flat}
  Let $p^\star$ be an {\em equilibrium price} that
  maximizes the revenue, i.e., $p^\star \in \arg \max_{p \in \set{P}} R(p).$
  The network is said to be {\em saturated} at $p$, if $\max_{t \in T} Y(t;p) =
  C_{\text{3g}}.$ For a {\bf \em unique} equilibrium price $p^{\star}$, the
  network is said to be {\em opt-saturated} if the network is saturated
  at $p^\star,$ and {\em opt-unsaturated} otherwise.
\end{definition}

Let $p_0$ be the {\em threshold price} above which all the feasible
prices lie, i.e., $p_0 = \inf_{p \in \set{P}} p.$
For a given price $p$, $R(p) > 0$ implies
$\max_{t \in T} Y(t;p)> 0,$ since no user subscription or no traffic
results in zero income to the provider. 
Note that $p^\star$ is not necessarily equal to $p_0.$ 
Let $A(p) = \max_{t \in T} Y(t;p)$ be the total 3G traffic at {\em peak} time.
Note that $A(p)$ is decreasing in $p.$
Since $A(p)$ is decreasing in $p,$ a feasible price in $\set{P}$ is greater
than or equal to $p_0.$




\subsection{Flat pricing}
\label{sec:analysis_flat}
This subsection considers the impact of offloading when flat pricing is used.
In flat pricing, 
a user pays a flat fee $p$ regardless of its 3G traffic
usage if it subscribes to the 3G service.  Since there is no incentive
to discourage excessive network traffic, 
the traffic volume
generated by a user equals to its traffic demand; $x_{\Phi}(t) =
\phi(t)$
for a subscribing user with total traffic demand $\Phi,$
where $\phi(t)$ is the traffic demand at slot $t$ split from $\Phi.$

A user with traffic demand $\Phi$ maximizes its net-utility:
\begin{eqnarray}
\sum_{t \in T} \gamma(t) x_{\Phi}(t)^{\theta} - p
   =  \sum_{t \in T} w(t)^{1-\theta} \phi(t)^{\theta} - p
  =  \Phi^{\theta} - p, 
  \label{eqn:user-flat}
\end{eqnarray}
since $\gamma(t) = w(t)^{1-\theta}$ by {\bf A3}, 
$x_\Phi(t) = \phi(t)$,
for all $t$, 
and the temporal preference $w(t) = \frac{\phi(t)}{\Phi}$. 
From \eqref{eq:revenue}, the provider maximizes its
revenue:
\ifthenelse{\boolean{singlever}}{
\begin{eqnarray}
R(p)  =
  \hat{N} p   \int_{p^{\frac{1}{\theta}}}^{\Phi_{\max}} \text{d} F_{\Phi}
  - \hat{N} \eta  \int_{p^{\frac{1}{\theta}}}^{\Phi_{\max}} \Phi  \text{d} F_{\Phi}
     =  \hat{N} \int_{p^{\frac{1}{\theta}}}^{\Phi_{\max}} (p-\eta\Phi) \text{d} F_{\Phi}, \label{eqn:provider-flat}
\end{eqnarray}
}
{
\begin{eqnarray}
R(p) & =  &
  \hat{N} p   \int_{p^{\frac{1}{\theta}}}^{\Phi_{\max}} \text{d} F_{\Phi}
  - \hat{N} \eta  \int_{p^{\frac{1}{\theta}}}^{\Phi_{\max}} \Phi  \text{d} F_{\Phi} \nonumber \\
    & = & \hat{N} \int_{p^{\frac{1}{\theta}}}^{\Phi_{\max}} (p-\eta\Phi) \text{d} F_{\Phi}, \label{eqn:provider-flat}
\end{eqnarray}
}
where $p^{1/\theta}$ is the lowest traffic demand of a subscriber 
(i.e., positive net-utility and thus $\Phi^{\theta} > p$).
No users subscribe if the price is too high, i.e., if $p
\geq p_{\max}$ from \eqref{eqn:user-flat}, where $p_{\max} =
\Phi_{\max}^{\theta}.$ Then, we should have
that $\set{P} \subset [0, p_{\max}).$ 

Our main results, Prop.~\ref{thm:unimodal_flat} and Theorem~\ref{thm:flat},
state how the economic values (e.g. price and revenue)
change by offloading. 
Prop.~\ref{thm:unimodal_flat} 
characterizes (i) $R(p)$ and the feasible price set, and
(ii) the equilibrium prices in the opt-saturated case
and (iii) the opt-unsaturated case.
Theorem~\ref{thm:flat} states that offloading is economically
beneficial for the users, the provider, and the regulator. 

\smallskip
\begin{proposition}[Equilibrium Price in Flat]
  \label{thm:unimodal_flat}
  If the cost coefficient $\eta<( \kappa_{\text{avg}} 
  {\Phi}_{\max}^{1-\theta} )^{-1},$ 
  \begin{enumerate}[(i)]
  \item $R(p)$ is unimodal\footnote{
	A function $f(x)$ is called unimodal, if for some value $v,$ 
	it is monotonically increasing for $x \leq v$ and monotonically
	decreasing for $x \geq v.$}	
   over $[0, p_{\max}),$ and the
    feasible price set $\set{P}$ is non-empty and connected.
  \item The network is {\em opt-saturated}, if $R'(p_{0}) < 0,$ where the
    unique equilibrium price $p^\star = p_0,$ where $p_0 =
    {\Phi}_{\max}^{\theta} \left( 1 - \frac{C_{3g}}{\kappa_{\text{peak}} \hat{N}
        \expect{{\Phi}}} \right) ^ {\frac{\theta}{2-\sigma}}.$
  \item The network is {\em opt-unsaturated}, if $R'(p_{0}) > 0,$ where
    the unique equilibrium price $p^\star = p^\star(\kappa_{\text{avg}})$ is such that
    $\frac{\partial R(p)}{\partial p}\big|_{p=p^\star} = 0,$ and $\frac{\partial p^\star(\kappa_{\text{avg}})}{\partial
      \kappa_{\text{avg}}} > 0.$
  \end{enumerate}
\end{proposition}

\smallskip
\begin{theorem}[Economic Gain from Offloading in Flat]
  \label{thm:flat}
  If $\eta < ( \kappa_{\text{avg}} {\Phi}_{\max}^{1-\theta} )^{-1},$ the net-utilities of
  all subscribers increase and the provider's revenue at equilibrium
  increases (thus the user surplus and the social welfare increase), as
  (i) $\kappa_{\text{peak}}$ decreases in the opt-saturated case, and (ii) as $\kappa_{\text{avg}}$ 
  decreases in the opt-unsaturated case.
\end{theorem}
\smallskip

\begin{figure}[t!]
    \center
    \includegraphics[width=1\columnwidth]{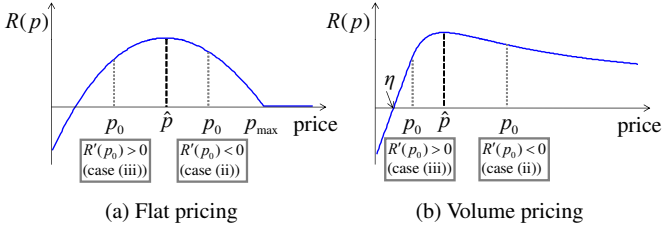}   
    \caption{
      Revenue function $R(p)$ in flat and volume pricing.
      The $p_{\max}$ is the highest price above which no user subscribes in flat pricing,
      $\hat{p}$ is the unique solution of $\frac{\partial R(p)}{\partial p} = 0,$
      and $\eta$ is the network cost coefficient.
      The achievable revenue (at equilibrium) is not always at $\hat{p}$,
      since $\hat{p}$ may not be in the feasible price set, which is determined by capacity constraint and provider rationality.
      \label{fig:analysis_revenue}}
\end{figure}

The proof is presented in the Appendix. Here, we briefly interpret
Prop.~\ref{thm:unimodal_flat} and Theorem~\ref{thm:flat}.  
First, clearly if the network cost is
too high, the provider cannot achieve any positive revenue, where the
condition of $\eta < ( \kappa_{\text{avg}} {\Phi}_{\max}^{1-\theta} )^{-1}$ guarantees the
existence of prices under which the revenue is positive. This condition
is relaxed as more offloading occurs (i.e., $\kappa_{\text{avg}}$ decreases), 
resulting in less restricted business condition 
with positive revenue from the
provider's perspective. 
Second, every feasible price is larger than or
equal to the threshold price $p_0.$ Also, $R(p)$ is unimodal
and $\set{P}$ is connected. 
Thus, 
at the threshold price, 
if $R'(p_0) < 0,$ 
then $R(p_0) \geq R(p)$ for all $p \in
\set{P}$ (see Fig.~\ref{fig:analysis_revenue}(a)).  Thus, the
equilibrium price is unique, and $p^\star = p_0,$ where $p_0$ is
characterized as in Prop.~\ref{thm:unimodal_flat}(ii). 
Also, 
the network is opt-saturated if $R'(p_0) < 0,$, because the peak traffic
volume $A(p_0) = C_{\text{3g}}$ (otherwise, there exists a smaller feasible
price than $p_0$).  Now, if $R'(p_0) > 0,$ the equilibrium price
$p^\star$ is such that $R'(p^\star)= 0,$ as in
Prop.~\ref{thm:unimodal_flat}(iii). Again, this case makes the network
opt-unsaturated because $A(p^\star) < A(p_0)$ (due to decreasing
property of $A(p)$ in $p$) and $A(p_0) \leq C_{\text{3g}}.$



 We now explain the relationship between traffic demand and
 opt-saturatedness.  The amount of total traffic demand of users
 affects the revenue change rate at the {\em traffic-maximizing} price,
 $R'(p_0),$ where 3G capacity and offloading indicators are fixed.  If
 traffic demand is high enough, when the provider can reduce its flat
 fee (by offloading or network upgrade),
 revenue increases even with the reduced price, i.e., $R'(p_0) <
 0,$ since increase of subscribers exceeds price reduction.  If traffic
 demand is not significantly high, subscription ratio does not increase
 drastically, so that revenue decreases, i.e., $R'(p_0) > 0,$ and at
 the optimal price, the 3G capacity is not fully utilized by the users.

Using the results of Prop.~\ref{thm:unimodal_flat},
Theorem~\ref{thm:flat} states that offloading is economically beneficial
from the perspective of the users, the provider, and the regulator,
where 
the mechanisms behind the increase in the revenue are different in the opt-saturated and opt-unsaturated cases.
In the opt-saturated case, as more 3G traffic is offloaded through WiFi
at {\em peak} time, i.e., $\kappa_{\text{peak}}$ decreases, 
the provider turns out to have
extra 3G capacity.  Then, the provider attracts more subscribers by
lowering its flat fee, in order to utilize the extra capacity.  As the
increase in the number of subscribers exceeds the reduced price, the
revenue increases. Indeed, from Prop.~\ref{thm:unimodal_flat}(ii)
the equilibrium price decreases as $\kappa_{\text{peak}}$ decreases.  
The net-utility
increases for all subscribers by price reduction since a subscribing
user in flat pricing 
always generate all the traffic demand. In the opt-unsaturated
case, the number of subscribers does not increase drastically
even if the provider decreases its flat fee, so that
the provider's income does not increase. However, the
network cost decreases substantially as the 3G traffic decreases, i.e.,
$\kappa_{\text{avg}}$ decreases, and the revenue increases.  
Since the equilibrium
price still decreases, as $\kappa_{\text{avg}}$ decreases from
Prop.~\ref{thm:unimodal_flat}(iii), the net-utility increases for
all subscribers
by the same argument 
as in the opt-saturated case.

\subsection{Volume pricing}
\label{sec:analysis_volume}

In volume pricing, user payment is proportional to its 3G traffic
volume.  For a given unit price, a user chooses the amount of traffic
that maximizes its net-utility. In this subsection, for tractability, we
focus more on the average analysis by assuming that per-user and -time 
dependence of delay profile and WiFi connection probability are homogeneous,
i.e., $\alpha_{\Phi}^{d} = \alpha^d, ~e_{\Phi}^d(t)= e^d$ for all $t$
and all users. Then, by \eqref{eqn:Y(t)}, for a user with traffic demand
$\Phi,$
\begin{eqnarray}
  \label{eq:sumy}
\sum_{t \in T} y_{\Phi}(t) \hspace{-0.05cm} = \hspace{-0.05cm} \sum_{t \in T} \Big ( x_{\Phi}(t)
\sum_{d=0}^{D} \alpha^d \big( 1 - e^d\big) \Big ) \hspace{-0.05cm} = \hspace{-0.05cm} \kappa_{\text{avg}} \sum_{t \in T} x_{\Phi}(t),
\end{eqnarray}
where note that $\kappa_{\text{avg}} = \sum_{d=0}^{D} \alpha^d \big( 1 - e^d\big)$ 
by the definition in \eqref{eq:def_q_and_k}.
A user with traffic demand $\Phi$
pays $p \sum_{t \in {T}} y_{\Phi}(t)$ and maximizes
the following net-utility (for a given price $p$):
\ifthenelse{\boolean{singlever}}{
\begin{eqnarray}
  \sum_{t \in T}  \gamma (t) {x}_{\Phi} (t)^{\theta} -
  p \sum_{t \in T}  {y}_{\Phi}(t)
  =    \sum_{t \in T} w (t)^{1-\theta} x_{\Phi} (t)^{\theta} -
  p q \sum_{t \in T}  {x}_{\Phi}(t)
  \label{eq:utility_volume}  
\end{eqnarray}
}
{
\begin{multline}
  \quad \quad
  \sum_{t \in T}  \gamma (t) {x}_{\Phi} (t)^{\theta} -
  p \sum_{t \in T}  {y}_{\Phi}(t) \cr
  \hspace{-0.5cm}=    \sum_{t \in T} w (t)^{1-\theta} x_{\Phi} (t)^{\theta} -
  p \kappa_{\text{avg}} \sum_{t \in T}  {x}_{\Phi}(t) \quad \quad \quad
  \label{eq:utility_volume}  
\end{multline}
}
using $\gamma(t) = w(t)^{1-\theta}$ and \eqref{eq:sumy}.
Since users pay in proportion to the volume of 3G traffic, 
a notion of {\em payment per unit (3G + WiFi) traffic}
is useful, given by:
\begin{eqnarray*}
  \frac{p \sum_{t \in T} y_{\Phi}(t)}{\sum_{t \in T} x_{\Phi}(t)}
  = \frac{p \kappa_{\text{avg}} \sum_{t \in T} x_{\Phi}(t)}{\sum_{t \in T} x_{\Phi}(t)}
  = p \kappa_{\text{avg}},
\end{eqnarray*}
where without offloading, i.e., $\kappa_{\text{avg}}=1,$ payment per unit traffic
is just $p.$ From \eqref{eq:revenue}, the provider maximizes its
revenue
\begin{eqnarray}
R(p) & = & \hat{N} (p - \eta)   \int_0^{\Phi_{\max}}  \sum_{t \in T} {y}_{\Phi}(t) \text{d} F_{\Phi}. \label{eqn:provider-volume}
\end{eqnarray}
Since $R(p) \leq 0$ for $p \leq \eta,$
we should have $\set{P} \subset (\eta, \infty).$
It is clear that as the payment per unit traffic $p \kappa_{\text{avg}}$ decreases, the
traffic which can be delivered per dollar increases. 
We will show that
the payment per unit traffic $p \kappa_{\text{avg}}$ decreases, as more offloading
through WiFi occurs. 

Similarly in flat pricing,
we present our main results by
Prop.~\ref{thm:unimodal_volume} and Theorem~\ref{thm:volume}.
Prop.~\ref{thm:unimodal_volume} characterizes
the equilibrium prices in opt-saturated
and opt-unsaturated cases.
Theorem~\ref{thm:volume} states that offloading is economically
beneficial for the users, the provider, and the regulator. 

\smallskip
\begin{proposition}[Equilibrium Price in Volume]
  \label{thm:unimodal_volume}
  \mbox{}
  \begin{enumerate}[(i)]
  \item $R(p)$ is unimodal for $p \geq \eta$ and 
    the feasible price set
    $\set{P}$ is non-empty and connected.
  \item The network is {\em opt-saturated}, if $R'(p_0) < 0,$
    where the unique equilibrium price is $p^\star = p_0$
      where $p_0 = p_0(\kappa_{\text{peak}})$ is the threshold price, and $\frac{\partial p_0(\kappa_{\text{peak}})}{\partial \kappa_{\text{peak}}} > 0.$
  \item
    The network is {\em opt-unsaturated}, if $R'(p_{0}) > 0,$ where
    the unique equilibrium price $p^\star = p^\star(\kappa_{\text{avg}})$ is such that
    $\frac{\partial R(p)}{\partial p}\big|_{p=p^\star} = 0,$ and
    $\frac{\partial (p^\star(\kappa_{\text{avg}}) \kappa_{\text{avg}})}{\partial
      \kappa_{\text{avg}}} > 0.$
  \end{enumerate}
\end{proposition}

\smallskip
\begin{theorem}[Economic Gain from Offloading in Volume]
  \label{thm:volume}
  The net-utilities for all users increase and the
  provider's revenue increases (thus the user surplus and the
  social welfare increase), (i) as $\kappa_{\text{peak}}$ decreases, in the opt-saturated
  case, and (ii) as $\kappa_{\text{avg}}$ 
  decreases, in the opt-unsaturated case.
\end{theorem}
\smallskip

The proof is presented in the Appendix.
Here, we briefly interpret Prop.~\ref{thm:unimodal_volume} and Theorem~\ref{thm:volume}.
Note that different from in flat pricing, in volume pricing, $\set{P}
\neq \emptyset$ for {\em any} network cost coefficient $\eta$ and all
users subscribe to the service. The revenue function is unimodal (see
Fig.~\ref{fig:analysis_revenue}(b)), so that our analysis becomes
significantly convenient, and as in flat pricing, $R'(p_0)$, determines
whether the threshold price $p_0$ is the price at equilibrium or not. 
The relationship between the traffic demand and opt-saturatedness
is analogous to that in flat pricing, where basically a high traffic
demand induces opt-saturatedness.
We also see that as the offloaded traffic increases, the payment per unit
traffic decreases in both opt-saturated and opt-unsaturated cases.  
In the opt-saturated case, we experience the revenue increase similarly
to flat pricing. In this case, offloading generates extra 3G capacity,
thereby the provider attracts more traffic by reducing the unit 
price 
to get higher revenue  
(Note that in flat, 
higher revenue is due to more subscribing users by
reducing the flat fee).
Indeed, from Prop.~\ref{thm:unimodal_volume}(ii), the equilibrium
price $p^\star$ decreases as $\kappa_{\text{peak}}$ decreases.
However, {\em in the opt-unsaturated case, flat and volume pricing behave
differently.}  Unlike the flat pricing, as 3G traffic is reduced by
offloading, the provider's income decreases in volume pricing. 
Then, the provider increases the price to compensate for the revenue decrease, which,
however, does not bring decrease in income for the following reasons:
even though the price $p$ is increased, 3G+WiFi traffic is not reduced
as long as the payment per unit traffic $p \kappa_{\text{avg}}$ is not 
increased.  Thus, the total income $p \kappa_{\text{avg}} \times$traffic 
(3G+WiFi) remains the same if the $p \kappa_{\text{avg}}$ is the same.  
Since cost is proportional to 3G traffic, reduced 3G
traffic leads to cost reduction which is the main factor to the
revenue increase. 
The payment per unit traffic at equilibrium
$p^\star(\kappa_{\text{avg}}) \kappa_{\text{avg}}$ decreases as 
$\kappa_{\text{avg}}$ decreases from Prop.~\ref{thm:unimodal_volume}(iii).

\section{Trace-driven numerical analysis}
\label{sec:numerical_results}

\subsection{Setup}

In this subsection, we describe the setup for our trace-driven numerical
analysis, such as the real traces and the parameter
values.  
The duration of a time slot is set to be an hour, i.e., $T = 24.$ The
number of BS cells is 31 and the average number of users per cell is 1000,\footnote{
This is a typical number of users in a macro BS.
For example, Sprint has 66,000 BSs and 55 million subscribers
at the end of 2011~\cite{number_of_bss,number_of_subscribers}.} 
thus total 31000 users. The choice of 31 cells is due to a real trace
which will be explained later.  We test two cellular capacities, 8 Mbps
for 3G and 32 Mbps for 4G, where the 4G capacity is projected to be
about four times the 3G capacity \cite{4g_capacity_gain}.\footnote{Yet,
  we still use the notation $C_{\text{3g}}$ for notational simplicity.}  We
set the price sensitivity $\theta=0.5$ and the cost coefficient $\eta =
0.1.$
We use two real traces to get the statistics of users' traffic and
WiFi connection probability.



\begin{figure}[h!]
  \centering
  \includegraphics*[width=0.95\columnwidth]{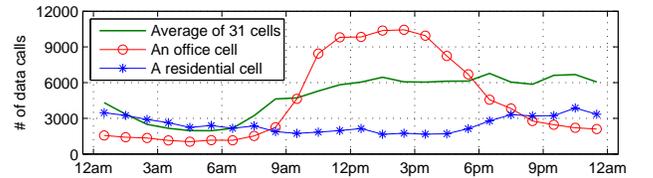} 
  \caption{
    The number of data calls on average and in office/residential cells.
    \label{fig:time_preference}}
\end{figure}

\smallskip
\noindent{\bf \em Trace 1 (3G traffic usage).}
The first trace is from a major cellular provider in Korea and includes
the information on the number of high speed downlink/uplink packet
access (HSDPA/HSUPA) calls, recorded every hour, in each of the 31 BSs in a
day.  Fig. \ref{fig:time_preference} shows 
the average number of calls per hour over 31 BSs, and 
that in an office and a residential cell, 
where we regard a cell as an office cell if the
call arrival at daytime (8:00 a.m. - 8:00 p.m.) exceeds that at night, and a
residential cell otherwise. 
There exist 15 office and 16 residential cells.
Assuming that the average data consumption per each call is similar over cells and time slots,
we regard the number of data calls as the amount of traffic demand at each cell and slot. 

\smallskip
\noindent{\bf \em Trace 2 (WiFi connection).}
The second trace is measured by 93 iPhone users from an iPhone user
community in Korea, who volunteer and record their time-varying WiFi
connectivity and locations, 
periodically scanned and
recorded at every 3 minutes for two weeks \cite{conext10}. 
Occupations of participants
were diverse, e.g. students, daytime workers, and freelancers,
as well as residential areas were,
where half of participants lived in Seoul.
We only recorded APs to which users can transmit data by sending a ping packet
to our server.
i.e., the trace only captures accessible WiFi APs that are open or users have authority. 

We now present the main parameters 
based on the {\em traces 1, 2} and the measurement results revealed 
in other research.

\smallskip
\noindent {\bf \em (a) Traffic demand (${\phi}_i$) and willingness to pay ($\gamma_i(t)$)}:
Most measurements on mobile
data~\cite{smartphone_usage:mobisys10,cellular_traffic,smartphone_traffic,cellular_traffic:sigmetrics11}
showed that the user traffic volume follows an upper-truncated power-law distribution
as used in the analysis of Section~\ref{sec:analysis}.
Thus, we use an upper-truncated power-law distribution\footnote{
  $f_{{\Phi}} (x) = {x^{-\sigma}}/{Z},$ for $0 \leq x \leq
  {\Phi}_{\max},$ where ${\Phi}_{\max}$ is the maximum value of ${\Phi}$
  and $Z = \frac{{\Phi}_{\max}^{1-\sigma}}{1-\sigma}.$ } 
  with exponent $\sigma = 0.57$ 
  (which is observed in \cite{cellular_traffic:sigmetrics11}),
  as the distribution of total daily traffic demand
${\Phi}_i$ by scaling the average, so that the per-month average ranges
from 93 MB to 5.2 GB. 
Note that in \cite{cisco12}, 1.3 GB/month is projected in year 2015.
The temporal preference ($w_i(t) = \phi_i(t)/\Phi_i$) 
of users follows the average temporal {\em usage pattern} in {\em trace 1}.  
Users' willingness to pay is set to include some randomness across users,
and its time-dependence is set to be proportional to temporal
preference, i.e., $\gamma_i(t) = \nu_i w(t)^{1-\theta},$ where $\nu_i$
is uniformly distributed over $(0, 1).$



\smallskip
\noindent {\bf \em (b) WiFi connection probability
  ($e_i^d(t)$)}: 
We use the {\em trace 2} to obtain the values of $(e_i^d(t): i \in N, t
\in
T).$ 
Since {\em trace 2} includes only 93 users, we repeatedly use their
individual traces to generate $N$ users' data, i.e., about $N/93$ users
have the same $e_i^d(t).$ We refer the readers to \cite{conext10} to
know how often users meet WiFi in the experiment. 
For 10 mins and 6 hours deadline, the average WiFi contact
probabilities are 0.7 and 0.88, and the medians 
are 0.87 and 0.97, resp.





\smallskip
\noindent {\bf \em (c) BS association ($s_i(t)$)}:
The information on cell-level mobility is important in our model due to
the (i) cell-level capacity constraint $C_{\text{3g}}$ 
(which requires to track the number of users in a cell) 
and (ii) congestion pricing (which
applies difference prices depending on spatio-temporal information).
Unfortunately, {\em trace 1} does not include users' cell-level mobility
mainly because of privacy, thus we combine {\em trace 1} with the
statistics from \cite{cellular_traffic} that has the distribution on the
number of distinct BSs visited by a user per day. 
At the first time slot,
we assign user associations so that  
the number of users in each cell at the 
first time slot is proportional to the traffic volume 
of the cell at that slot (e.g. 8 a.m.) in {\em trace 1}.
We assign handover probabilities (from the associated cell to other cells)
to users 
so that (i) the expected number of users in each cell 
at the next time slot
is proportional to the traffic volume in {\em trace 1} and
(ii) the number of visited BSs follows the statistics in~\cite{cellular_traffic},
assuming that handovers are uniformly distributed among time slots. 

\smallskip
\noindent {\bf \em (d) Delay profile
  ($\alpha_{i}^d$)}: 
To model the delay profile of
users, 
we use various scenarios, such as {\em no-deadline}, {\em short}, {\em
  medium}, and {\em long}, where each scenario consists of four
different classes (Video, Data, P2P, and Audio), as classified by Cisco
\cite{cisco12}.  The details are described in
Table~\ref{tlb:traffic_classification}. 
We consider the economic impact
of WiFi offloading for two cases in our numerical results: (i) all users
are uniformly given a scenario, e.g., {\em medium}, and (ii) there is
a fixed portion of users for each scenario.




\begin{table}[h!]
  \centering
  \caption{
    Traffic classification projected in year 2015 from Cisco~\cite{cisco12}
    and assigned deadlines for each traffic class. SC: Scenario
  }
  \label{tlb:traffic_classification}
  \tabcolsep 3pt
    \begin{tabular}{|c|c|c|c|c|c|}
    \hline
    & Video & Data & P2P & Audio (VoIP) & Total   \\
    \hline \hline
    Ratio & 66.4 \% & 20.9 \% & 6.1 \% & 6.6 \% & 100 \% \\   
    \hline
    SC:zero   & 0 sec. & 0 sec. & 0 sec. & 0 sec. & - \\
    SC:short   & 10 min. & 30 min. & 10 min. & 0 sec. & - \\
    SC:medium   & 30 min. & 1 hour & 30 min. & 0 sec. & - \\
    SC:long & 2 hours & 6 hours & 2 hours & 0 sec. & -  \\
    \hline
  \end{tabular}
\end{table}

\subsection{Results}
\label{sec:results}

Our numerical results quantify the benefits of delayed WiFi offloading in
various aspects. We present our results by summarizing the key
observations.

%

\begin{figure}[t!]
  \centering
  \subfigure[$C_{\text{3g}}$ = 3.6 GB/hour (8 Mbps)]
  {\includegraphics*[width=0.517\columnwidth]{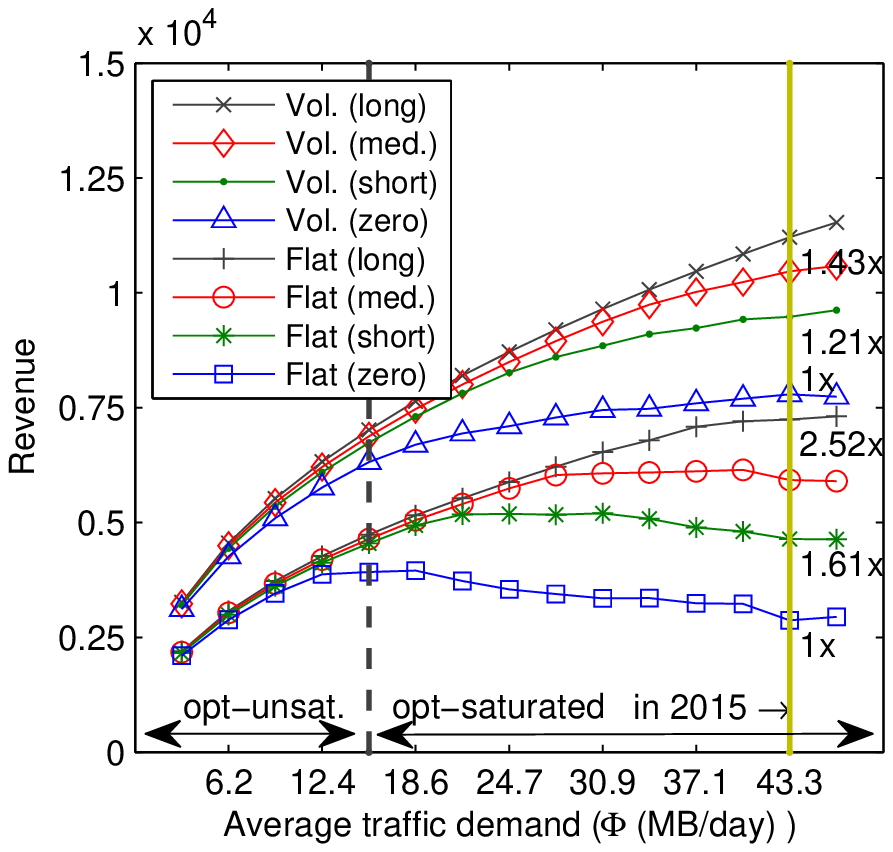}}
  \hspace{-0.2cm}
  \subfigure[$C_{\text{3g}}$ = 14.4 GB/hour (32 Mbps)]
  {\includegraphics*[width=0.481\columnwidth]{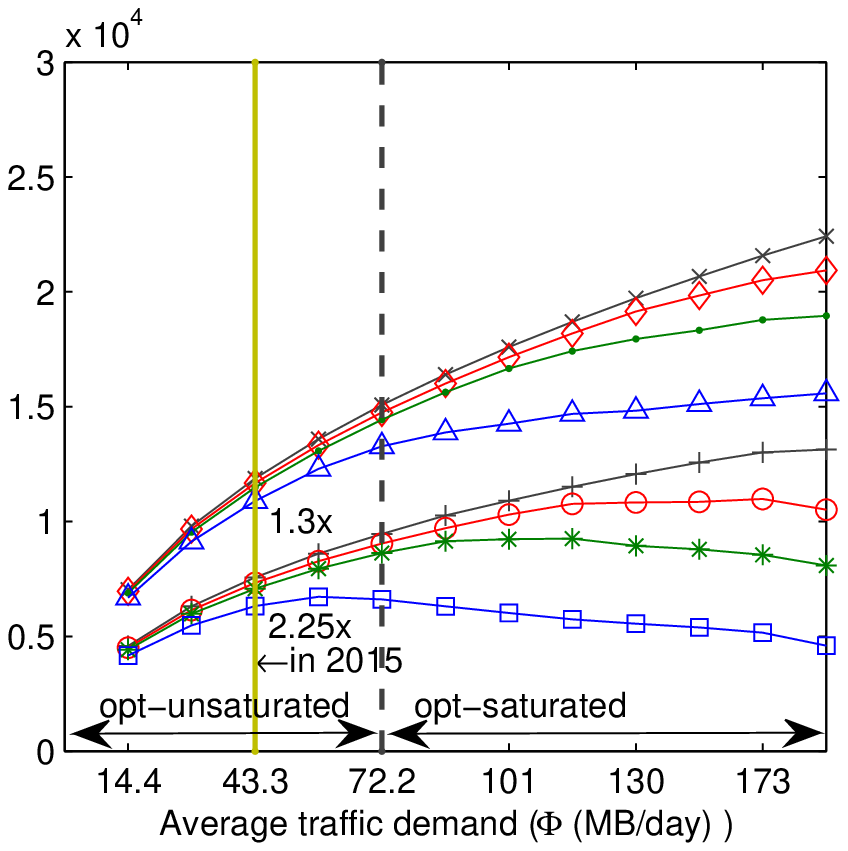}}
  \caption{Flat and volume pricing: revenue for various delay profiles in Table~\ref{tlb:traffic_classification} and traffic demand, with different cellular capacities,
  where users experience a single scenario in Table~\ref{tlb:traffic_classification}.
  Delay profiles are set to be the same across all users.
  Offloading indicators are decreasing as delay tolerance gets higher (from short to long),
  where $\kappa_{\text{avg}} = .44,~.28,~.23,~.15$ and $\kappa_{\text{peak}} = .0044,~.0026,~.0020,~.0013$ for zero, short, medium, and long scenarios.
  The numbers (-x) represent the increase of revenue 
  by (a) delayed offloading or (b) network upgrade (from 3G to 4G).
  Opt-saturatedness and opt-unsaturatedness are determined by the traffic demand and cellular capacity,
  where the dotted line shows the threshold in the on-the-spot offloading over flat pricing.
    \label{fig:flat_volume_3g}
  }
\end{figure}

\begin{figure}[t!]
  \center
  \includegraphics[width=0.8\columnwidth]{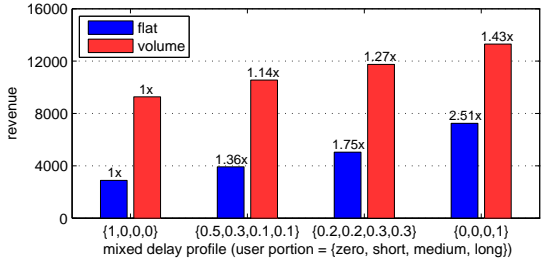}     
  \caption{
    Revenue in flat and volume pricing with various mixed delay profiles,
    where user portions of scenarios (zero, short, medium, long) are different.
    The average traffic demand is 43.3 MB/day (1.5 GB/month) and 3G capacity is 3.6 GB/hour.
    The numbers (-x) represent the increase of revenue compared to
    the revenue in on-the-spot offloading.
    \label{fig:mixed_delay_profile}}
\end{figure}

\smallskip
\noindent{\bf \em 1) Revenue in volume pricing exceeds that in
  flat pricing by applying delayed WiFi offloading, but the revenue
  increase is higher in flat pricing than volume pricing:}
Fig. \ref{fig:flat_volume_3g}(a) depicts the revenue of flat and
volume for various traffic demand and delay profiles,
where users experience a single scenario.  
Revenue in volume exceeds that in flat in all cases, because
in flat, a subscriber with high traffic demand generates heavy
traffic and dominates the network resources without paying more fees to
the provider, whereas in volume,
user payment is proportional to traffic volume,
so that if a subscriber generates heavy traffic, the payment is high.
This imposes \emph{negative externality} (i.e., congestion) 
to the provider and reduces provider revenue.
However, the revenue {\em increase} of delayed offloading,
which is the amount of increased revenue over revenue in
the on-the-spot offloading, is higher in flat. 
The revenue increase in flat pricing is about
61-152\%, 
whereas the revenue increase in volume pricing is about 21-43\%, when the
average traffic demand is 43.3 MB/day (1.5 GB/month).  
This is because in flat pricing,
3G traffic reduction does not affect the provider's income, whereas, in
volume pricing, 3G traffic reduction decreases the income.  
We also
depict the revenue of flat and volume pricing when users have a mixture
of four scenarios of delay deadline in
Fig.~\ref{fig:mixed_delay_profile}.  
We find that as user portion with
high delay tolerance increases, revenue increases in both flat and
volume pricing.

\smallskip
\noindent{\bf \em 2)
The revenue gain from on-the-spot to delayed offloading is similar to
that generated by the network upgrade from 3G to 4G:}
Fig. \ref{fig:flat_volume_3g} depicts revenue of flat and volume
pricing for various traffic demand, with different cellular capacities.
As the traffic demand increases, the network gets congested and eventually
becomes opt-saturated.
If a provider upgrades the cellular network, such as a 4G network, 
the revenue increases by 115\% in flat and 30\% in volume, respectively, 
when the traffic demand is 43.3 MB/day.  We recall
that the revenue gain of offloading in flat and volume pricing is 61-152\%
and 21-43\%, which is as significant as the revenue gain from
network upgrade from 3G to 4G.
Thus, if traffic demand is high compared to capacity,
adopting delayed offloading can be a good solution to increase revenue,
where the network upgrade induces huge installation costs.
Note that when traffic demand is not high (i.e., opt-unsaturated),
the revenue increase is small both in network upgrade and delayed offloading. 

\begin{figure}[t!]
  \centering
  \subfigure[Flat: fee and subscription ratio]
  {
  \includegraphics*[width=0.234\textwidth]{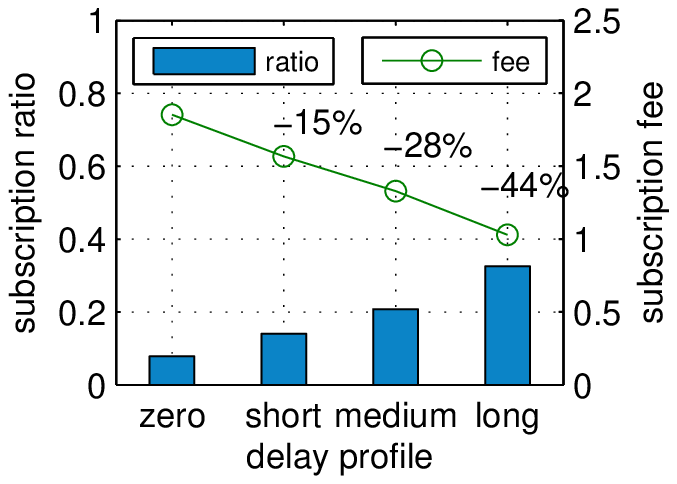}  
  \label{fig:flat_fee_ratio}
  }
  \subfigure[Volume: payment per unit traffic]
  {
  \includegraphics*[width=0.211\textwidth]{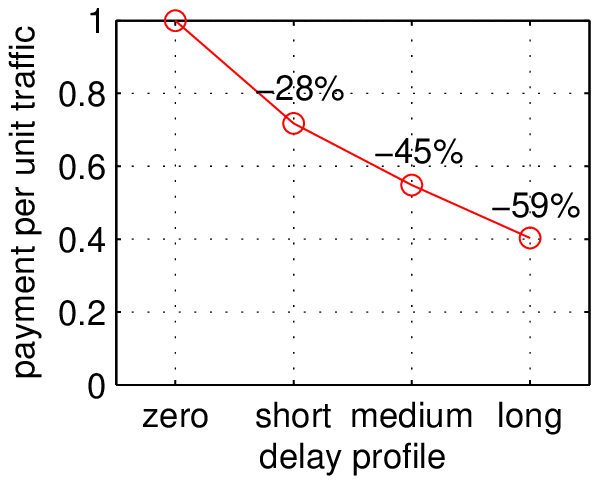}    
  \label{fig:volume price}
  }
  \caption{
    Change of flat price and subscription ratio in flat pricing,
    and payment per unit traffic in volume pricing. 
    The average traffic demand is 43.3 MB/day (1.5GB/month) and 3G capacity is 3.6 GB/hour (opt-saturated).
  }
  \label{fig:price_flat_volume}
\end{figure}

%

\begin{figure}[t!]
  \centering
  \subfigure[Flat vs. Two-tier]
  {
  \hspace{-0.2cm}
  \includegraphics*[width=0.49\columnwidth]{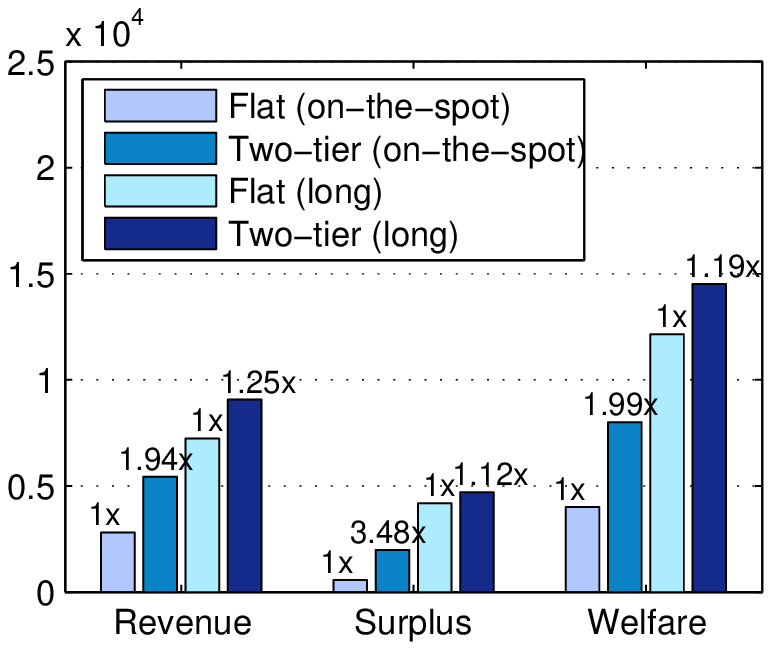}
  \label{fig:tiered_vs_flat}
  \hspace{-0.3cm}
  }
  \subfigure[Volume vs. Congestion]
  {
  \hspace{-0.3cm}
  \includegraphics*[width=0.49\columnwidth]{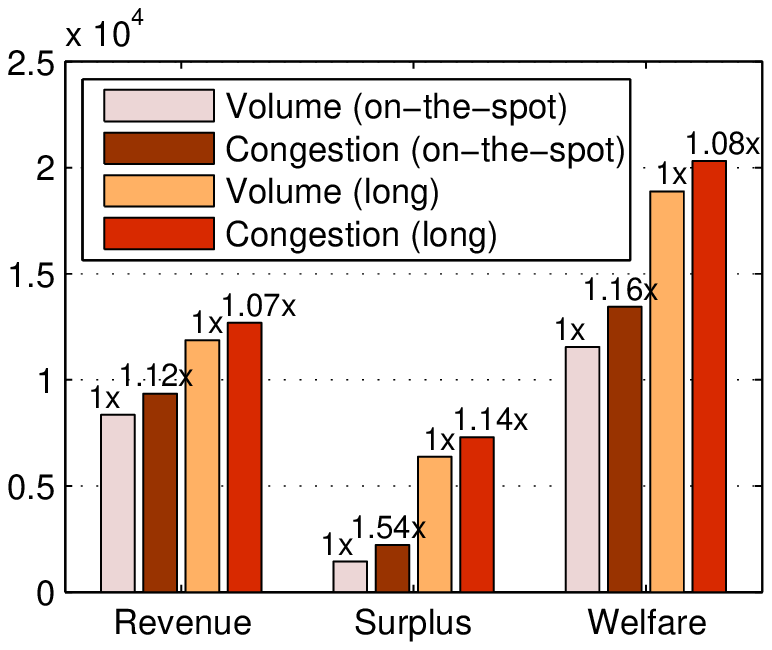}
  \label{fig:congestion_vs_volume}
  \hspace{-0.2cm}  
  }
  \caption{
    The revenue, surplus, and welfare in flat, tiered, volume and congestion pricing.
    The number (-x) above each bar represents the increase compared to the revenue in (a) flat or (b) volume pricing.
    The average traffic demand is set to be 43.3MB/day (1.5GB/month) which is
    projected in year 2015 by Cisco
    and the 3G capacity is 3.6GB/hour.
    \label{fig:tiered_and_congestion}
  }
\end{figure}

\smallskip
\noindent{\bf \em 3)
As more traffic is offloaded, the flat price decreases 
  and subscription ratio increases simultaneously 
  in flat pricing,
  and payment per unit traffic decreases in volume pricing:}
As shown in Fig.~\ref{fig:price_flat_volume},
the flat fee decreases by 15-44\% 
and subscription ratio increases accordingly
in flat pricing,
and payment per unit traffic decreases by 28-59\% in volume pricing,
when the average traffic demand is 1.5 GB/month (opt-saturated).
In the opt-unsaturated case, price reduction is not drastic
both in flat and volume pricing, since the {\em income} does not increase by
price reduction due to low traffic demand.
The reduction in flat price and payment per unit traffic induces increase
in user surplus, as shown in Fig.~\ref{fig:tiered_and_congestion}.

\begin{figure}[t!]
  \center
  \includegraphics[width=0.7\columnwidth]{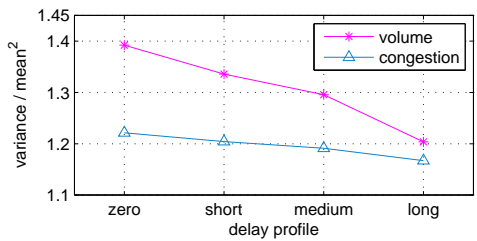}     
  \caption{
    Variance of normalized cell load at each time slot.
    \label{fig:variance_congestion}}
\end{figure}

\smallskip
\noindent{\bf \em 4) Two-tier and congestion pricing
  increase the revenue, compared to flat and volume pricing, but such
  gains become smaller, as more traffic is offloaded:} 
Fig. \ref{fig:tiered_and_congestion} shows
the change of revenue, surplus, and welfare in four pricing schemes.
It is intuitive that as pricing granularity increases in terms of price
(from flat to tiered) or space/time (from volume to congestion),
revenue increases, because the provider has more
degree of freedom to control the market.  However, the rate of increase 
diminishes as more traffic is offloaded through WiFi.  
Using the traffic demand
in 2015, the revenue in two-tier pricing is greater than that in flat pricing
by 94\% and 25\% in on-the-spot and delayed offloading,
resp., where the revenue in congestion pricing is greater
than that in volume pricing by 12\% and 7\% in on-the-spot and delayed
offloading, resp. 
We also find that delayed offloading reduces spatiotemporal
imbalance by dispersing traffic to other time and locations, so that the
effect of space/time-varying price is reduced.  To see this, we depict
the variance of normalized cell load at each time in
Fig. \ref{fig:variance_congestion}, where the variance decreases as more
traffic is offloaded.
As a result, the time-flattening effect and revenue increase of 
congestion pricing decrease,
since congestion pricing performs better when 
temporal imbalance in traffic demand is severe. 





\begin{figure}[t!]
  \centering
  \subfigure[Decrease in the revenue gain.]
  {
  \includegraphics[width=0.46\columnwidth]{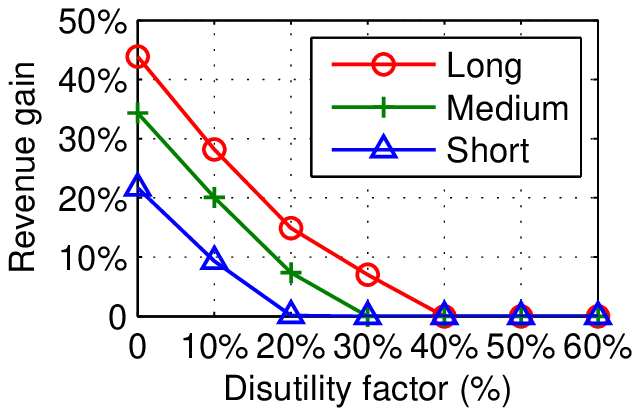}     
  }
\subfigure[Decrease in the fraction of delayed offloading users.]
{
  \includegraphics[width=0.46\columnwidth]{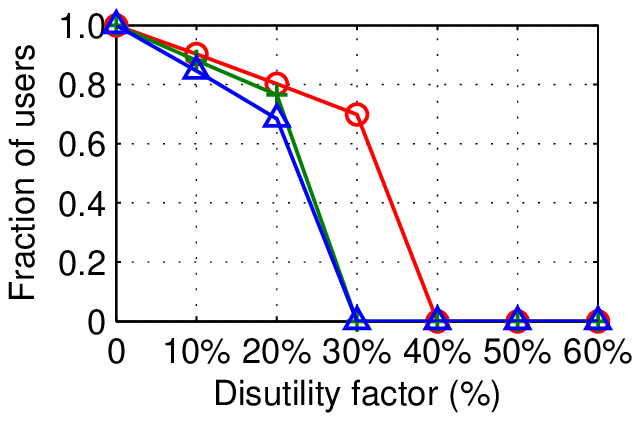}     
  }
    \caption{Effect of delay disutility on (a) the revenue gain and (b) the fraction of delayed offloading users}
    \label{fig:disutility_volume}
\end{figure}


\smallskip
\noindent{\bf \em 5)
As users' utility decreases by the delay disutility,
the revenue gain from delayed offloading decreases,
but the decrease in revenue gain is not severe
when users have high delay tolerance:} 
Here, we try to understand the effect of delay disutility on revenue gain
by applying the {\em disutility factor} in users' utility.
We consider the volume pricing and
assume that utility of users who perform delayed offloading is decreased by
the {\em disutility factor} (\%) from the original utility without delayed offloading.
A user can decide whether it adopts delayed offloading or not,
based on the net-utility,
i.e., only if the price discount is higher than the amount of disutility,
the user adopts delayed offloading.
We depict the revenue gain and the fraction of users who adopt delayed offloading, 
in Fig.~\ref{fig:disutility_volume}.
Both the revenue gain and the delayed offloading ratio decrease as the {\em disutility factor} increases.
We still obtain more than 50\% of the revenue gain 
when users have less than 15\% and 10\% of disutility factor,
for the {\em long} and {\em short} delay profiles, resp.
Note that users will experience less delay disutilty for shorter delay.
When the disutility factor is higher than 40\%, 
no user decides to use delayed offloading, so that the revenue gain is zero.

\section{Concluding Remark and Future Work}
\label{sec:conclusion}


In this paper, we model a game-theoretic framework to study the economic
aspects of WiFi offloading, where we drew the following messages from
the analytical and numerical studies: First, WiFi offloading is
economically beneficial for both a monopoly provider and users, where
the economic gains are not ignorable. Also, a simple pricing is enough
in the sense that two-tier and volume-based pricing do not increase the
revenue and the net-utilities for higher offloading chances, which is
true as of now and in the future, when more WiFi APs are expected to be
deployed. 

Another well-known benefit from WiFi offloading is 
Smartphone energy reduction,
which is shown in~\cite{delayenergy:mobisys10,conext10,IMC_energy09}.
This is mainly because short range communication (e.g. WiFi) 
typically has lower energy-per-bit
than long range communication (e.g. 3G/4G).
It is shown that 50-60\% of transmission energy can be reduced for 1-hour delay~\cite{conext10}.
Even though we do not consider the energy benefit in this paper,
it is obvious that most users benefit from the increased battery lifetime.


There are a few limitations in our work. Our results rely on the
assumption that network traffic has a degree of delay tolerance and
users can tolerate some amount of delay, where delay depends on the
class of traffic 
(even if we provide the numerical results of a mixture
of users with different delay scenarios, 
including ones requiring no delay deadline). 
Thus, our results can sometimes be regarded as an
upper-bound on the economic benefits of WiFi offloading.
A simple way
of reflecting those limitations would be to design a net-utility
function which jointly captures the happiness by data transmission and
the disutility by delay, which we leave as a future work. 
Another future work is to consider multiple providers,
where they have different plans to overcome the mobile data explosion
(e.g. delayed offloading, network upgrade, and complex pricing).

\bibliographystyle{IEEEtran}
\bibliography{Reference}

\section{Appendix}
\label{sec:appendix}

\subsection{Proof of Proposition \ref{thm:unimodal_flat}}




\noindent\underline{\em Step 1: Unimodality of $R(p)$.}
In flat pricing, the total 3G traffic volume of a 3G-subscribing user is
equal to its total traffic demand, i.e., $\sum_{t \in T} x_{\Phi}(t) =
\sum_{t \in T} \phi(t) = \Phi,$ and the user enters the market only when $\Phi
> p^{\frac{1}{\theta} },$ from \eqref{eqn:user-flat}. Based on it, we
first express $\sum_{x \in T} X(t)$ and $A(p)$ using the
parameters of the traffic demand distribution. We first have:
  \begin{eqnarray*}
    \sum_{t \in T} X(t)  = \hat{N} \int_{p^{\frac{1}{\theta}}}^{{\Phi}_{\max}}  {\Phi} f_{\Phi} ({\Phi}) d \Phi
    = \hat{N} \frac{ {\Phi}_{\max}^{2-\sigma} - p^{\frac{2-\sigma}{\theta}}}{Z (2-\sigma)},
  \end{eqnarray*}
where $f_{\Phi} ( {\Phi}) = \frac{ {\Phi}^{-\sigma}}{Z} $ by {\bf A2}
and recall that $Z = \frac{{\Phi}_{\max}^{1-\sigma}}{1-\sigma}.$
Then, by \eqref{eq:def_q_and_k},
\begin{eqnarray}
  \label{eq:max_A}
  A(p) = \kappa_{\text{peak}} \sum_{t \in T} X(t)
  =  \kappa_{\text{peak}} \hat{N} \frac{ {\Phi}_{\max}^{2-\sigma} 
  - p^{\frac{2-\sigma}{\theta}}}{Z (2-\sigma)}.
\end{eqnarray}
From \eqref{eqn:provider-flat}, the revenue $R(p)$ is expressed as:
\ifthenelse{\boolean{singlever}}{
  \begin{eqnarray}
    \label{eq:revenue_flat}
    R(p) 
    =  \hat{N} p \left( 1 - \frac{p^{\frac{1-\sigma}{\theta}}}{Z(1-\sigma)} \right)    
    - \frac{\eta \kappa_{\text{avg}} \hat{N} \left(  {\Phi}_{\max}^{2-\sigma} 
    - p^{\frac{2-\sigma}{\theta}} \right) }{Z (2-\sigma)}. \nonumber
  \end{eqnarray}
}
{
  \begin{eqnarray}
    \label{eq:revenue_flat}
    R(p) 
    & = &  \hat{N} p \left( 1 - \frac{p^{\frac{1-\sigma}{\theta}}}{Z(1-\sigma)} \right)
      - \frac{\eta \kappa_{\text{avg}} \hat{N} \left(  {\Phi}_{\max}^{2-\sigma} 
      - p^{\frac{2-\sigma}{\theta}} \right) }{Z (2-\sigma)}. \nonumber
  \end{eqnarray}
}
Then, the first derivative of $R(p)$ is
\begin{eqnarray}
  \frac{\partial R(p)}{\partial p} = \hat{N} \left( 1 - \frac{\left( 1 + \frac{1-\sigma}{\theta} \right) p^{\frac{1-\sigma}{\theta}}}{Z(1-\sigma)}
    + \frac{\eta \kappa_{\text{avg}} p^{\frac{2-\sigma}{\theta}-1}}{Z \theta} \right).  \label{eq:objective_first}
\end{eqnarray}

We can easily check that $R'(0) >0$ and $R'(p_{\max}) <0$ under the
condition that $\eta < ( \kappa_{\text{avg}} {\Phi}_{\max}^{1-\theta} )^{-1}$.
and from the
intermediate value theorem, there exists a price $\hat{p} \in (0,
p_{\max})$ such that $R'(\hat{p}) = 0.$
Now, we show that $R(p)$ is unimodal, thereby $\hat{p}$ is unique.
The second derivative of $R(p)$ is
\begin{eqnarray}
  \frac{\partial^2 R(p)}{{\partial p}^2}
  \hspace{-0.05cm} = \hspace{-0.05cm} 
  \hat{N} p^{\frac{1-\sigma-\theta}{\theta}} 
  \frac{\eta \kappa_{\text{avg}} p^{\frac{1-\theta}{\theta}} ( 2 - \theta - \sigma ) 
  \hspace{-0.05cm} - \hspace{-0.05cm} (1+\theta-\sigma) }{Z \theta^2}.
  \label{eq:objective_second}
\end{eqnarray}
We have ${\partial^2 R(p)}/{{\partial p}^2} <0$ (concave) over $[0, \bar{p}),$
and ${\partial^2 R(p)}/{{\partial p}^2} >0$ (convex) over $(\bar{p},p_{\max}),$
where $\bar{p} = \left( \frac{1+\theta-\sigma}{\eta \kappa_{\text{avg}} (2-\theta-\sigma)} \right)^{\frac{\theta}{1-\theta}},$
such that ${\partial^2 R(p)}/{{\partial p}^2} = 0.$
Since $R'(p_{\max}) < 0$ and $R'(p)$ is increasing over $(\bar{p},p_{\max})$ from the convexity,
$R'(p) < 0$ over $(\bar{p},p_{\max}),$ and $R'(\bar{p}) < 0.$
Thus, $\hat{p} \notin [\bar{p},p_{\max}).$
Since $R'(0) > 0,$ $R'(\bar{p}) < 0,$ and $R'(p)$ is decreasing over $[0, \bar{p})$ from the concavity,
the solution of $R'(\hat{p}) = 0,$ $\hat{p}$ is unique and $\hat{p} < \bar{p}.$
Also, $R(p)$ is unimodal over $[0, p_{\max})$ since $R'(p)$ has only one sign change. 

\smallskip
\noindent\underline{\em Step 2: Characterization of $\set{P}$.}
By definition of the set of feasible prices with provider rationality
and capacity constraint, $\set{P} = \set{E} \cap \set{F}$ where $
\set{E} = \{ p ~|~ A(p) \leq C_{3g}\}$ and $\set{F} = \{ p ~|~ R(p) >0
\}$.
From \eqref{eq:max_A}, $A(p)$ is decreasing in $p.$ Thus,
there exists some
$p_{\min},$ such that $A( p) \leq C_{3g} \Leftrightarrow p \geq
p_{\min}.$ Therefore, $\set{E} = \{ p ~|~ p \geq p_{\min} \}$ which is a
connected set. The $p_{\min}$ is characterized as:
\begin{eqnarray*}
  p_{\min} = \left\{ \begin{array}{ll}
      0 & \mbox{if $A(0) < C_{\text{3g}}$} \\
      \left(  {\Phi}_{\max}^{2-\sigma} -
        \frac{C_{3g} Z (2-\sigma)}{  \hat{N} \kappa_{\text{peak}}} \right)^{\frac{\theta}{2-\sigma}}
      & \mbox{if $A(0) \geq C_{\text{3g}}$}
    \end{array}
  \right.
\end{eqnarray*}
Note that $p_{\min} < \Phi^{\theta}_{\max}$ since $0 < \sigma < 1.$
Regarding $\set{F}$, we first recall that $p < p_{\max} = \Phi_{\max}^{\theta}$ since $R(p) = 0$
for $p \geq {\Phi}^{\theta}_{\max}$ (i.e., there exists no subscriber).
Since $R(p)$ is unimodal over $ [0, p_{\max}),$
$\set{F} = \{ p ~|~ R(p) > 0\}$ is connected.
There exists a unique $p_{z} < \hat{p}$ such that $R(p_{z}) = 0,$
since $R(0) \leq 0, ~R(\hat{p}) > 0 $ and $R(p)$ is strictly increasing ($R'(p) >0$) in $0 \leq p \leq \hat{p}.$
Hence $\set{F} = (p_{z}, ~ p_{\max})$ for $p_{z} < p_{\max}$ such that $R(p_{z}) = 0.$
Since both $\set{E}$ and $\set{F}$ are connected, $\set{P} = \set{E}
\cap \set{F} $ is connected. Note that $p_0 = \inf \{ p | R(p) > 0,
A(p) \leq C_{\text{3g}} \} = \max\{p_{\min},p_z\}.$ If $R(p_0) > 0$,
$p_0 \in \set{F}$ and $\set{P} = [p_0, ~ p_{\max}).$ If $R(p_0) = 0,$
then, $p_0 \notin \set{F}$. Thus, $p_0 = p_{z}$ and $\set{P} = (p_0, ~
p_{\max}).$ From {\em Steps 1} and {\em 2,} the result (i) holds.


\smallskip
\noindent\underline{\em Step 3: Proof of (ii).}
If $R'(p_0) < 0$, then $\hat{p} < p_0,$ which means $\hat{p}$ is not in
$\set{P}$, i.e, $\hat{p}$ cannot be the equilibrium price.
Since $R(p_{\max}) = 0$ and $R'(p) < 0$ for $p_0 \leq p \leq p_{\max}$ from unimodality of $R(p)$
we have $R(p_0) > 0$; $\set{P} = [p_0 ~p_{\max}).$   
Therefore $p_0 = p_{\min}$ such that $A(p_{\min})
= C_{\text{3g}}$ and
\ifthenelse{\boolean{singlever}}{
  \begin{eqnarray*}
    p_0
    = \left(  {\Phi}_{\max}^{2-\sigma} -
      \frac{C_{3g} Z (2-\sigma)}{  \hat{N} k} \right)^{\frac{\theta}{2-\sigma}}
    = {\Phi}_{\max}^{\theta} \left( 1 - \frac{C_{3g}}{\hat{N} k \expect{{\Phi}}} \right) ^ {\frac{\theta}{2-\sigma}},
  \end{eqnarray*}
}
{
  \begin{eqnarray*}
    p_0
    & = & \left(  {\Phi}_{\max}^{2-\sigma} -
      \frac{C_{3g} Z (2-\sigma)}{\kappa_{\text{peak}} \hat{N}} \right)^{\frac{\theta}{2-\sigma}} \\
    & = & {\Phi}_{\max}^{\theta} \left( 1 - \frac{C_{3g}}{\kappa_{\text{peak}} \hat{N} \expect{{\Phi}}} \right) ^ {\frac{\theta}{2-\sigma}},
  \end{eqnarray*}
}
where $\expect{\Phi} = \frac{1-\sigma}{2-\sigma} \Phi_{\max}.$
Since $R(p)$ is decreasing over $\set{P}$,
$R(p)$ is maximum at $p_0$, that is, $p_0$ is the unique equilibrium price.
Since $A(p_0)= C_{\text{3g}}$, the network is {\em opt-saturated}.

\smallskip
\noindent\underline{\em Step 4: Proof of (iii).}
If $R'(p_0) > 0$, then $\hat{p} > p_0$ since $R(p)$ is unimodal. 
From \eqref{eq:max_A}, $A(p)$ is a decreasing function
of $p$. Hence $A(\hat{p}) < A(p_0) \leq C_{3g}$; the network is {\em opt-unsaturated}.
We will show that the derivative $\frac{\partial \hat{p}}{ \partial \kappa_{\text{avg}}}$ 
is positive.
From \eqref{eq:objective_first} and taking a derivative of $\kappa_{\text{avg}}$ 
w.r.t. $\hat{p},$
\begin{eqnarray*}
  \kappa_{\text{avg}} = \frac{\frac{1+\theta-\sigma}{1-\sigma} 
  \hat{p}^{\frac{1-\sigma}{\theta}} - Z \theta }
  {\eta \hat{p}^{\frac{2-\sigma}{\theta}-1}}, \quad  
  \frac{\partial \kappa_{\text{avg}}}{\partial \hat{p}} = \frac{\hat{p}^{\frac{-2+\sigma}{\theta}}}{\eta (1-\sigma)} g(\hat{p}^{\frac{1-\sigma}{\theta}}), \label{eq:dq_dp}
\end{eqnarray*}
where $  g(z) = (2-\theta-\sigma) {\Phi}_{\max}^{1-\sigma} -
  \frac{(1-\theta)(1+\theta-\sigma)}{\theta} z.$
To show $\frac{\partial \hat{p}}{ \partial \kappa_{\text{avg}}} > 0$, 
it suffices to show
$\frac{\partial \kappa_{\text{avg}}}{\partial \hat{p}} > 0$, for which we show
$g(\hat{p}^{\frac{1-\sigma}{\theta}}) > 0$.  From
\eqref{eq:objective_first}, we have \ifthenelse{\boolean{singlever}}{
  \begin{eqnarray*}
    \hat{p}^{\frac{1-\sigma}{\theta}}
    =  \frac{\theta {\Phi}_{\max}^{1-\sigma}}
    {(1+\theta-\sigma) - \eta q (1-\sigma) \hat{p}^{\frac{1-\theta}{\theta}}}
    < \frac{\theta {\Phi}_{\max}^{1-\sigma}}{(1+\theta-\sigma) \left( 1 - \frac{1-\sigma}{2-\theta-\sigma}\right) }
    = \frac{\theta(2-\theta-\sigma)}{(1-\theta)(1+\theta-\sigma)} {\Phi}_{\max}^{1-\sigma},
  \end{eqnarray*}
}
{
  \begin{eqnarray*}
    \hat{p}^{\frac{1-\sigma}{\theta}}
    & = & \frac{\theta {\Phi}_{\max}^{1-\sigma}}
    {(1+\theta-\sigma) - \eta \kappa_{\text{avg}} (1-\sigma) \hat{p}^{\frac{1-\theta}{\theta}}}
    \cr
    & < & \frac{\theta {\Phi}_{\max}^{1-\sigma}}{(1+\theta-\sigma) \left( 1 - \frac{1-\sigma}{2-\theta-\sigma}\right) } \cr
    & = & \frac{\theta(2-\theta-\sigma)}{(1-\theta)(1+\theta-\sigma)} {\Phi}_{\max}^{1-\sigma},
  \end{eqnarray*}
}
since $\hat{p} < \bar{p},$ where $\bar{p} = \left( \frac{1+\theta-\sigma}{\eta \kappa_{\text{avg}} (2-\theta-\sigma)} \right)^{\frac{\theta}{1-\theta}}$
such that ${\partial^2 R(p)}/{{\partial p}^2} = 0$ from \eqref{eq:objective_second}.
Let $\rho =  \frac{\theta(2-\theta-\sigma)}{(1-\theta)(1+\theta-\sigma)} {\Phi}_{\max}^{1-\sigma}$.
Since $g(z)$ is decreasing as $z$ is increasing and $ g(\rho) = 0$,
  $g(\hat{p}^{\frac{1-\sigma}{\theta}})>   g(\rho) = 0$
since $\theta \in (0,1)$ and $\sigma \in (0,1).$
Thus, $\frac{\partial \kappa_{\text{avg}}}{\partial \hat{p}} > 0 $
and $\frac{\partial \hat{p}}{\partial \kappa_{\text{avg}}}$ is also positive. \QED

\begin{figure}[t!]
    \center
    \includegraphics[width=0.75\columnwidth]{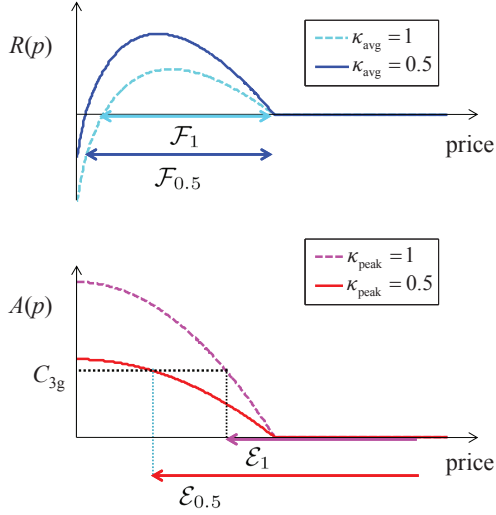}
    \caption{
      Change of feasible price set, $\set{P} = E \cap F$ for $\kappa_{\text{avg}} = 1,~ \kappa_{\text{peak}} = 1$ and $\kappa_{\text{avg}} = 0.5, ~ \kappa_{\text{peak}} = 0.5$ in flat pricing.
      \label{fig:flat_revenue_proof}}
\end{figure}


\subsection{Proof of Theorem \ref{thm:flat}}

\noindent \underline{\em (i) Opt-saturated.} 
Recall that
net-utility of a subscriber with ${\Phi}$ is ${\Phi}^{\theta}-p,$ where
${\Phi} = \sum_{t \in T} {\phi}(t).$ 
Hence, the net-utility of a
subscriber increases as $p^\star$ decreases from
Proposition~\ref{thm:unimodal_flat}(ii).

To study the impact of
$\kappa_{\text{peak}}$, we regard $\kappa_{\text{peak}}$ as a variable, not a constant.  
To show that  $R(p^\star)$ increases as $\kappa_{\text{peak}}$ decreases,
we will show that  $ \frac{\partial R(p^\star)}{\partial \kappa_{\text{peak}}} <0$.
The first derivative of $R(p^{\star})$ with respect to $\kappa_{\text{peak}}$ is,
\begin{eqnarray*}
\frac{\partial R(p^\star)}{\partial \kappa_{\text{peak}}}
  =  \frac{\partial R(p^\star)}{\partial p^\star}  \frac{\partial p^\star}{\partial \kappa_{\text{peak}}}.
\end{eqnarray*}
From Proposition~\ref{thm:unimodal_flat}(ii), 
$p^{\star} $ is an increasing function in $\kappa_{\text{peak}}$.
Therefore,  it suffices to show
that $ \frac{\partial R(p^\star)}{\partial p^\star} < 0$, which is
proved in the proof of Proposition~\ref{thm:unimodal_flat}(ii). 


\smallskip
\noindent \underline{\em (ii) Opt-unsaturated.}  Using a similar
argument in the proof of Theorem~\ref{thm:flat}(i), the net-utility of a
user and user surplus increase as $\kappa_{\text{avg}}$ decreases, 
since the flat fee decreases as $\kappa_{\text{avg}}$ decreases from
Proposition~\ref{thm:unimodal_flat}(iii). 
In terms of the provider's revenue, we use the notations $R(p,\kappa_{\text{avg}})$ and
$\set{P}_{\kappa_{\text{avg}}}$ 
to explicitly express the dependence of $R(p)$ and $\set{P}$ on
$\kappa_{\text{avg}},$ 
because our interest lies in examining how $R(p)$ and $\set{P}$ change with varying
$\kappa_{\text{avg}}.$ 
First, $R(p,\kappa_{\text{avg}})$ increases 
as $\kappa_{\text{avg}}$ decreases for $p \in [0, p_{\max}],$ 
since differentiating by $\kappa_{\text{avg}}$ yields:
\begin{eqnarray}
  \frac{\partial R(p,\kappa_{\text{avg}})}{\partial \kappa_{\text{avg}}} =
  - \frac{\eta \hat{N} \left( \Phi_{\max}^{2-\sigma} - p^{\frac{2-\sigma}{\theta}} \right) }{Z (2-\sigma)} < 0.
\end{eqnarray}
Second, we will show that the set $\set{P}_{\kappa_{\text{avg}}}$ 
gets ``enlarged'' as $\kappa_{\text{avg}}$ 
decreases, 
for which it suffices to show that revenue increases.
$\set{P}_{\kappa_{\text{avg}}} 
= \set{E}
\cap \set{F}_{\kappa_{\text{avg}}},$ where 
$\set{E} = \{p ~ | ~ p \geq p_{\min}\}$ and
$\set{F}_{\kappa_{\text{avg}}} = \{p ~ | ~ R(p,\kappa_{\text{avg}}) > 0 \}.$
Note that $\set{E}$ does not depend on $\kappa_{\text{avg}},$ but on $\kappa_{\text{peak}}.$
Since $R(p,\kappa_{\text{avg}})$ is decreasing in $\kappa_{\text{avg}}$,
$\set{F}_{\kappa_{\text{avg}}}$ is enlarged as $\kappa_{\text{avg}}$ decreases.
Therefore, $\set{P}_{\kappa_{\text{avg}}}$ 
is enlarged (or remains the same) and 
$\max_{p \in \set{P}_{\kappa_{\text{avg}}}} 
R(p,\kappa_{\text{avg}})$ increases as $\kappa_{\text{avg}}$ 
decreases.
We illustrate this in Fig. \ref{fig:flat_revenue_proof}.
\QED



\subsection{Proof of Proposition \ref{thm:unimodal_volume}}

\noindent\underline{\em Step 1: Unimodality of $R(p)$.}
  From \eqref{eq:utility_volume}, 
  \begin{eqnarray}
  \label{eq:net_utility_volume}
    {U}(\bm{x}_{\Phi})
    =  \sum_{t \in {T}} \left( w (t)^{1-\theta} x_{\Phi} (t)^{\theta} 
    - p \kappa_{\text{avg}} x_{\Phi} (t) \right).
  \end{eqnarray}
  Let $v(x_{\Phi} (t)) =   w (t)^{1-\theta} x_{\Phi} (t)^{\theta} 
  - p \kappa_{\text{avg}} x_{\Phi} (t)$.
  The net-utility  $ {U}(\bm{x}_{\Phi})$ is maximized when  $v(x_{\Phi} (t))$
  is maximized for each $t$.
  At given $t$, 
  \ifthenelse{\boolean{singlever}}{
    \begin{eqnarray*}
      \frac{\partial v(x_{\Phi}(t))}{\partial x_{\Phi} (t)}
      = \theta w(t)^{1-\theta} x_{\Phi} (t)^{\theta-1} - p q, \quad
      \frac{\partial^2 v(x (t))}{{\partial x_{\Phi} (t)}^2}
      = - \theta (1-\theta) w(t)^{1-\theta} x_{\Phi} (t)^{\theta-2}.
    \end{eqnarray*}
  }
  {
    \begin{eqnarray*}
      \frac{\partial v(x_{\Phi}(t))}{\partial x_{\Phi} (t)}
      = \theta w(t)^{1-\theta} x_{\Phi} (t)^{\theta-1} - p \kappa_{\text{avg}},
    \end{eqnarray*}
    \begin{eqnarray*}
      \frac{\partial^2 v(x_{\Phi} (t))}{{\partial x_{\Phi} (t)}^2}
      = - \theta (1-\theta) w(t)^{1-\theta} x_{\Phi} (t)^{\theta-2}.
    \end{eqnarray*}
  }
  Since $\theta \in (0,1),$ $ \frac{\partial^2 v(x_{\Phi} (t))}{{\partial x_{\Phi} (t)}^2}<0 $.
  Therefore,  $v(x (t))$ is  concave in $x(t)$.
  Thus, at each time $t$, $v(x_{{\Phi}} (t))$ takes a unique maximum at
  \ifthenelse{\boolean{singlever}}{
    \begin{eqnarray*}
      {x}_{{\Phi}}^\star (t) = \min \left\{ {\phi} (t) , w(t) \left(\frac{\theta}{p q}\right)^{\frac{1}{1-\theta}}  \right\}
      = w(t)  \min  \left\{ \Phi,  \left(\frac{\theta}{p q}\right)^{\frac{1}{1-\theta}}  \right \} .
    \end{eqnarray*}
  }
  {
    \begin{eqnarray*}
      {x}_{{\Phi}}^\star (t) & =  & \min \left\{ {\phi} (t) , w(t) \left(\frac{\theta}
      {p \kappa_{\text{avg}}}\right)^{\frac{1}{1-\theta}}  \right\}  \\
      & = &w(t)  \min  \left\{ \Phi,  \left(\frac{\theta}
      {p \kappa_{\text{avg}}}\right)^{\frac{1}{1-\theta}}  \right \} .
    \end{eqnarray*}
  }
  The second equality holds since  $\phi(t) = w(t) \Phi$  and $\Phi = \sum_{t \in {T}} \phi(t).$
  Moreover, it can be easily shown that
  $ v(x^{\star}_{{\Phi}} (t))>0$
  for all $t$.  This  implies that a user with positive $\Phi$ subscribes the service and $x_{\Phi}^{\star}(t) \neq 0$.
  Therefore,
  \begin{eqnarray*}
    \sum_{t \in {T}} {x}_{{\Phi}}^\star (t)
    = \left \{ \begin{array}{ll}
        \Phi & \mbox{ if $\Phi < \left(\frac{\theta}
        {p \kappa_{\text{avg}}}\right)^{\frac{1}{1-\theta}}$} \\
        \Psi(p)
        & \mbox{ if $\Phi > \left(\frac{\theta}
        {p \kappa_{\text{avg}}}\right)^{\frac{1}{1-\theta}}$}
      \end{array}    \right.
  \end{eqnarray*}
  where  $\Psi(p) = \left(\frac{\theta}{p \kappa_{\text{avg}}}\right)^{\frac{1}{1-\theta}}.$
  Then, total user traffic over a day, $\sum_{t \in {T}} X(t)$ is as follows:
  \ifthenelse{\boolean{singlever}}{
    \begin{eqnarray*}
      \sum_{t \in {T}} X(t)
      & = & \hat{N} \int_{0}^{\Phi_{\max}} \sum_{t \in {T}} x^{\star}_{\phi}(t) f_{\Phi} ({\Phi}) d \Phi
        =  \hat{N} \left( \int_{0}^{\Psi(p)}  {\Phi} f_{\Phi} ({\Phi}) d \Phi + \int_{\Psi(p)  }^{ {\Phi}_{\max}} \Psi(p) f_{\Phi} ({\Phi}) d \Phi \right) \\
      & = & \left \{ \begin{array}{ll}
          \hat{N} \Psi(p) \left( 1 - \frac{\Psi(p)^{1-\sigma}}{(2-\sigma)\Phi_{\max}^{1-\sigma}} \right)
          & \mbox{ if $\Psi(p) \leq \Phi_{\max}. $ } \\
          \hat{N} \expect{\Phi}    & \mbox{ if $\Psi(p) > \Phi_{\max}. $}
        \end{array} \right.
    \end{eqnarray*}
  }
  {
    \begin{eqnarray*}
      \lefteqn{ \sum_{t \in {T}} X(t)
        =  \hat{N} \int_{0}^{\Phi_{\max}} \sum_{t \in {T}} x^{\star}_{\phi}(t) f_{\Phi} ({\Phi}) d \Phi }  \\
      & =& \hat{N} \left( \int_{0}^{\Psi(p)}  {\Phi} f_{\Phi} ({\Phi}) d \Phi + \int_{\Psi(p)  }^{ {\Phi}_{\max}} \Psi(p) f_{\Phi} ({\Phi}) d \Phi \right) \\
      &  = & \left \{ \begin{array}{ll}
          \hat{N} \Psi(p) \left( 1 - \frac{\Psi(p)^{1-\sigma}}{(2-\sigma)\Phi_{\max}^{1-\sigma}} \right)
          & \mbox{ if $\Psi(p) \leq \Phi_{\max}. $ } \\
          \hat{N} \expect{\Phi}    & \mbox{ if $\Psi(p) > \Phi_{\max}. $}
        \end{array} \right.
    \end{eqnarray*}
  }
  We denote
  \ifthenelse{\boolean{singlever}}{
    \begin{eqnarray}
      B(p)  = \sum_{t \in {T}} Y(t;p) = q \sum_{t \in {T}} X(t;p), \quad
      A(p)  = \max_{t \in {T}} Y(t;p) = k \sum_{t \in {T}} X(t;p).  \label{eq:max_A_volume}
    \end{eqnarray}
  }
  {
    \begin{eqnarray}
      B(p)  & = & \sum_{t \in {T}} Y(t;p) = \kappa_{\text{avg}} \sum_{t \in {T}} X(t;p) \nonumber \\
      A(p)  & = & \max_{t \in {T}} Y(t;p) = \kappa_{\text{peak}} \sum_{t \in {T}} X(t;p).  \label{eq:max_A_volume}
    \end{eqnarray}
  }
  By (\ref{eqn:provider-volume}),   the revenue of the provider is
  $R(p) =  (p-\eta) B(p).$

    %

  We now show that $R(p)$ is unimodal over $\set{P}.$
    Note that
    \begin{eqnarray*}
      \Psi(p) > {\Phi}_{\max} ~~ \Leftrightarrow ~~
      p < \theta \big( \kappa_{\text{avg}} {\Phi}_{\max}^{1-\theta} \big) ^ {-1}.
    \end{eqnarray*}
    If $p < \theta \big( \kappa_{\text{avg}} {\Phi}_{\max}^{1-\theta} \big) ^ {-1}$ ,
    then  $B(p)$ is a positive constant, $B(p) = \kappa_{\text{avg}} \hat{N} \expect{\Phi}$,
    for all $p \in \set{P}$     and
    $\frac{\partial R(p)}{\partial p} = \kappa_{\text{avg}} \hat{N} \expect{\Phi}$.
    We now consider the case
    \begin{eqnarray*}
      p \geq \theta \big( \kappa_{\text{avg}} {\Phi}_{\max}^{1-\theta} \big) ^ {-1} 
      ~~ \Leftrightarrow ~~
      \Psi(p) \leq {\Phi}_{\max}.
    \end{eqnarray*}
    Then  for  $  p \geq \theta \big( q {\Phi}_{\max}^{1-\theta} \big) ^ {-1},$
    \begin{eqnarray*}
      B(p) = \hat{N} \kappa_{\text{avg}} \Psi(p) \left( 1 - \frac{\Psi(p)^{1-\sigma}}{(2-\sigma)\Phi_{\max}^{1-\sigma}} \right) > 0,
    \end{eqnarray*}
    since  $B(p) >0$, $sgn(R'(p)) =  sgn{ \left( \frac{R'(p)}{B(p)} \right)} $.
    We will investigate $sgn(R'(p))$ by investigating
    $sgn{ \left( \frac{R'(p)}{B(p)} \right)} $ and
    show that $\frac{\partial R(p)}{\partial p}$ has a unique solution $\hat{p}$
    of  $\frac{\partial R(p)}{\partial p} = 0 $  by showing that
    ${ \frac{R'(p)}{B(p)}}  =0 $ has a unique solution at $\hat{p}$.
    The first and second derivatives of $B(p)$ are,
  \ifthenelse{\boolean{singlever}}{
    \begin{eqnarray*}
      \frac{\partial B(p)}{\partial p}  = \frac{-\hat{N} q \Psi(p)}{p(1-\theta)} \left( 1 - \frac{\Psi(p)^{1-\sigma}}{\Phi_{\max}^{1-\sigma}} \right), \quad
      \frac{\partial B(p)^2}{\partial^2 p} = \frac{ \hat{N} q \Psi(p)}{p^2 (1-\theta)^2}
      \left( (2-\theta) - \frac{(3-\sigma-\theta) \Psi(p)^{1-\sigma}}{\Phi_{\max}^{1-\sigma}} \right).
    \end{eqnarray*}
  }
  {
    \begin{eqnarray*}
      \frac{\partial B(p)}{\partial p}  
      & = & \frac{-\hat{N} \kappa_{\text{avg}} \Psi(p)}{p(1-\theta)} 
      \left( 1 - \frac{\Psi(p)^{1-\sigma}}{\Phi_{\max}^{1-\sigma}} \right), \cr
      \frac{\partial B(p)^2}{\partial^2 p} 
      & = & \frac{ \hat{N} \kappa_{\text{avg}} \Psi(p)}{p^2 (1-\theta)^2}
      \left( (2-\theta) - \frac{(3-\sigma-\theta) \Psi(p)^{1-\sigma}}{\Phi_{\max}^{1-\sigma}} \right).
    \end{eqnarray*}
  }
    Let $B'(p) = \frac{\partial B(p)}{\partial p} $ and $B''(p) = \frac{\partial B(p)^2}{\partial^2 p}$.
    The first derivative of revenue function $R(p)$ is
    \begin{eqnarray}
      \label{eq:objective_volume}
      \frac{\partial R(p)}{\partial p} =
      (p - \eta) B'(p) + B(p).
    \end{eqnarray}
    We have
  \ifthenelse{\boolean{singlever}}{
    \begin{eqnarray*}
      \frac{\partial \left( \frac{R'(p)}{B(p)} \right)}{\partial p}  
      = \frac{B'(p)}{B(p)} + (p - \eta) \frac{B''(p)B(p) - B'(p)^2}{B(p)^2}
      =  \frac{ - l(p) }{p^2 (1-\theta)^2 B(p)^2}  < 0,
    \end{eqnarray*}
  }
  {
    \begin{eqnarray*}
      \frac{\partial \left( \frac{R'(p)}{B(p)} \right)}{\partial p}  
      & = & \frac{B'(p)}{B(p)} + (p - \eta) \frac{B''(p)B(p) - B'(p)^2}{B(p)^2} \\
      & = &  \frac{ - l(p) }{p^2 (1-\theta)^2 B(p)^2}  < 0,
    \end{eqnarray*}
  }
    where
  \ifthenelse{\boolean{singlever}}{
    \begin{eqnarray*}
      l(p) = (p-\eta) \left( \frac{(1-\sigma)^2 \Psi(p)^{1-\sigma}}{(2-\sigma)\Phi_{\max}^{1-\sigma}} \right)  +
             \eta(1-\theta)\left(1-\frac{\Psi(p)^{1-\sigma}}{\Phi_{\max}^{1-\sigma}}\right)B(p)
      > 0.
    \end{eqnarray*}
  }
  {
    \begin{eqnarray*}
      l(p) & = & (p-\eta) \left( \frac{(1-\sigma)^2 \Psi(p)^{1-\sigma}}{(2-\sigma)\Phi_{\max}^{1-\sigma}} \right)  +  \\
      & & ~~~ \eta(1-\theta)\left(1-\frac{\Psi(p)^{1-\sigma}}{\Phi_{\max}^{1-\sigma}}\right)B(p) \\
      & > & 0.
    \end{eqnarray*}
  }
    since $p \geq \eta, ~0 \leq \Psi(p) \leq \Phi_{\max}, ~B(p) > 0,$ and $\theta \in (0,1).$
    Thus, $\frac{R'(p)}{B(p)}$ is strictly decreasing in $p.$
    Note that $\frac{R'(\eta)}{B(\eta)} = 1 > 0,$ and
  \ifthenelse{\boolean{singlever}}{
    \begin{eqnarray*}
      \lim_{p \rightarrow \infty} \frac{R'(p)}{B(p)}
      = 1 + \lim_{p \rightarrow \infty} \frac{- \frac{p-\eta}{p(1-\theta)}
        \left(1-\frac{\Psi(p)^{1-\sigma}}{\Phi_{\max}^{1-\sigma}}\right) }{1-\frac{\Psi(p)^{1-\sigma}}{(2-\sigma)\Phi_{\max}^{1-\sigma}}}
      = 1 - \frac{1}{1-\theta} < 0,
    \end{eqnarray*}
  }
  {
    \begin{eqnarray*}
      \lim_{p \rightarrow \infty} \frac{R'(p)}{B(p)}
      & = & 1 + \lim_{p \rightarrow \infty} \frac{- \frac{p-\eta}{p(1-\theta)}
        \left(1-\frac{\Psi(p)^{1-\sigma}}{\Phi_{\max}^{1-\sigma}}\right) }{1-\frac{\Psi(p)^{1-\sigma}}{(2-\sigma)\Phi_{\max}^{1-\sigma}}} \cr
      & = & 1 - \frac{1}{1-\theta} < 0,
    \end{eqnarray*}
  }
    since $\lim_{p \rightarrow \infty} \Psi(p) = 0$ and $\theta \in (0,1).$
    Therefore $\frac{R'(p)}{B(p} < 0$ for sufficiently large $p$.
    Considering that $ \frac{R'(p)}{B(p)}$ is a decreasing function of $p$ and
    $\frac{R'(\eta)}{B(\eta)} > 0$, there should be a unique $\hat{p} \in (\eta, \infty)$
    such that $  \frac{R'(p)}{B(p)} =0$ (note that $\hat{p} > \eta$). Since $B(p) >0$ for any $p$,
    $R'(p)=0$ only at $\hat{p}$. Summarizing, the sign of $R'(p)$ is
  \ifthenelse{\boolean{singlever}}{
    \begin{eqnarray*}
      sgn(R'(p) ) & = &  sgn { \left( \frac{R'(p)}{B(p)} \right)}
      = \left \{  \begin{array} {ll}
          + & \mbox{if $\eta \leq p < \hat{p}$} \\
          0 & \mbox{if $p = \hat{p} $} \\
          - & \mbox{if $p > \hat{p}$}.
        \end{array} \right.
    \end{eqnarray*}
  }
  {
    \begin{eqnarray*}
      sgn(R'(p) ) & = &  sgn { \left( \frac{R'(p)}{B(p)} \right)} \\
      & = & \left \{  \begin{array} {ll}
          + & \mbox{if $\eta \leq p < \hat{p}$} \\
          0 & \mbox{if $p = \hat{p} $} \\
          - & \mbox{if $p > \hat{p}$}.
        \end{array} \right.
    \end{eqnarray*}
  }
    Thus, $R(p)$ is unimodal for $p > \eta.$

\smallskip
\noindent\underline{\em Step 2: Characterization of $\set{P}$.}
    We now find the set of feasible prices  $\set{P}$.
    Since $\sum_{t \in {T}} X(t) >0$ and $\kappa_{\text{avg}} \leq 1$,
    the provider's revenue is negative if $p \leq \eta$
    from \eqref{eqn:provider-volume}.
    By the {\em provider rationality}, i.e., $R(p) > 0$ for $p \in \set{P},$
    \begin{eqnarray}
      p \in \set{P}  \Rightarrow  p > \eta. \label{eqn:E_volume}
    \end{eqnarray}

    $A(p)$ is nonincreasing in $p,$ since
    \begin{eqnarray*}
      \frac{\partial A(p)}{\partial p}
      = \left\{ \begin{array}{ll}
          0 & \mbox{ if $\Psi(p) > \Phi_{\max}$} \\
          \frac{-\hat{N} k \Psi(p)}{p(1-\theta)} \left( 1 - \frac{\Psi(p)^{1-\sigma}}{\Phi_{\max}^{1-\sigma}} \right)
          & \mbox{ if $\Psi(p) \leq \Phi_{\max}$}
        \end{array} \right.
    \end{eqnarray*}
    Thus, if $p \in \set{P}$, then
    \begin{eqnarray*}
      A( p) \leq C_{3g}  & \Leftrightarrow &     p ~ \geq ~ p_{\min},
    \end{eqnarray*}
    where $p_{\min} = \inf \{p ~|~ A(p) \leq C_{\text{3g}} \}.$
    Note that $p_0 = \max\{p_{\min},\eta\}.$
    Thus, if $\eta \geq p_{\min},$ $p_0 = \eta$ and $\set{P} = (p_0,\infty),$
    and if $\eta  < p_{\min}$, $p_0 = p_{\min}$ and $\set{P} = [p_0,\infty).$
    Note that $p_{\min} = 0$ if and only if
    $\kappa_{\text{peak}} \hat{N} \expect{\Phi} \leq C_{\text{3g}},$
    since $A(0) = \kappa_{\text{peak}} \hat{N} \expect{\Phi}.$
    From {\em Steps 1} and {\em 2,} the result (i) holds.

\begin{figure}[t!]
      \center
      \includegraphics[width=0.75\columnwidth]{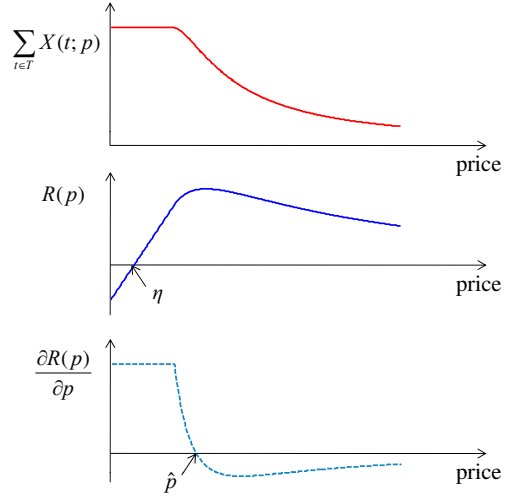} 
      \caption{
        An illustration of $\sum_{t \in T} X(t;p), ~ R(p),$ and $\frac{\partial R(p)}{\partial p}$ over volume price $p.$
        \label{fig:volume_proof}}
\end{figure}

\smallskip
\noindent\underline{\em Step 3: Proof of (ii).}
    If $R'(p_0) \leq 0,$ $\hat{p} \leq p_0,$ $R(p_0) > 0$ and $p_0$ is the unique optimal price,
    since $R'(p) \leq 0$ for $p \in \set{P}.$
    From $R(p_0) > 0,$ $p_0 > \eta \geq 0.$
    Thus, $A(p_0) = C_{3g}$ and the network is {\em opt-saturated}.

    We now show that 
    $\frac{\partial p_0(\kappa_{\text{peak}})}{\partial \kappa_{\text{peak}}} > 0.$
      To study the impact of $\kappa_{\text{peak}}$, 
      we regard $\kappa_{\text{peak}}$ as a variable, not a constant.
  Recall that   $p^\star = p_0$ and $A(p_0) = C_{\text{3g}}$ 
  if the network is {\em opt-saturated}.
    From \eqref{eq:def_q_and_k},
    \begin{eqnarray}
      \kappa_{\text{peak}} = \frac{C_{\text{3g}} }{\sum_{t \in {T}} X(t;p_0)}, 
      \label{eqn:k-alpha-vol-optsat}
    \end{eqnarray}
    where $X(t;p)$ is the total traffic arrival at time $t$ when price is $p.$
    We denote $V(p) = \sum_{t \in {T}} X(t;p).$
    Differentiating by $\kappa_{\text{peak}},$ we yield
    \begin{eqnarray}
      \label{eq:dpdk-volume}
      \frac{\partial p_0}{\partial \kappa_{\text{peak}}}
      & = & \frac{- V(p_0)^2}{C_{\text{3g}} V'(p_0)} > 0,
    \end{eqnarray}
    where
    \begin{eqnarray*}
      \frac{\partial V(p)}{\partial p}
      = \left\{ \begin{array}{ll}
          0 & \mbox{ if $\Psi(p) > \Phi_{\max}$} \\
          \frac{-\hat{N} \Psi(p)}{p(1-\theta)} \left( 1 - \frac{\Psi(p)^{1-\sigma}}{\Phi_{\max}^{1-\sigma}} \right)
          & \mbox{ if $\Psi(p) \leq \Phi_{\max}$}
        \end{array} \right.
    \end{eqnarray*}
    since $V(p) > 0$ and $V'(p) < 0,$
    for $p \in \set{P}$ in opt-saturated case.
    Note that $V'(p) = 0$ only if 
    $\kappa_{\text{peak}} V(p) = \kappa_{\text{peak}} N \expect{\Phi} > C_{\text{3g}}$
    and $\kappa_{\text{peak}} V(p) \leq C_{\text{3g}}$ 
    over $\set{P}$ in opt-saturated case.
   By (\ref{eq:dpdk-volume}), the optimal price $p^\star$ decreases and
    actual payment per unit traffic $p^\star \kappa_{\text{avg}}$ decreases 
    as $\kappa_{\text{peak}}$ decreases.
    We have proven (ii) holds.

\smallskip
\noindent\underline{\em Step 4: Proof of (iii).}
    We now consider the case $R'(p_0) > 0.$
    If $\eta \geq p_{\min},$ then $\set{P} = (\eta,\infty)$ and $\hat{p} > \eta$.
    Thus, $\hat{p}$ is the unique optimal price.
    To show that $A(\hat{p}) < C_{\text{3g}},$
    we consider two cases $p_{\min} = 0$ and $p_{\min} > 0.$
    Recall that $p_{\min} = 0$ if and only if
    $\kappa_{\text{peak}} \hat{N} \expect{\Phi} \leq C_{\text{3g}},$
    since $A(0) = \kappa_{\text{peak}} \hat{N} \expect{\Phi}.$
    Thus, if $p_{\min} = 0,$ 
    $A(p_{\min}) = \kappa_{\text{peak}} \hat{N} \expect{\Phi} \leq C_{\text{3g}}.$
    Note that $A'(p) < 0$ for $p$ such that $\Psi(p) \leq \Phi_{\max},$
    i.e., $p \geq \theta \big( \kappa_{\text{avg}} \Phi_{\max}^{1-\theta} \big).$
    Since $\hat{p} \geq \theta \big( \kappa_{\text{avg}} \Phi_{\max}^{1-\theta} \big),$
    $A'(\hat{p}) < 0$ and $A(\hat{p}) < A(p_{\min}) \leq C_{\text{3g}}.$
    If $p_{\min} > 0,$ 
    $A(p_{\min}) = C_{\text{3g}} < \kappa_{\text{peak}} \hat{N} \expect{\Phi}.$
    Since $A'(\hat{p}) < 0,$
    $A(\hat{p}) < A(p_0) \leq C_{\text{3g}}.$
    The network is {\em opt-unsaturated}.
    If $\eta < p_{\min},$ then $p_0 = p_{\min},$ $\set{P} = [p_{\min},\infty)$ 
    and $p_{\min} > 0.$
    Note that $A(p_{\min}) = C_{\text{3g}}$ if $p_{\min} > 0.$
    Recall that if $R'(p_0) > 0,$ 
    then $\hat{p} > p_0$ and $\hat{p}$ is the unique optimal price.
    Since $A(\hat{p}) < A(p_{\min}) =  C_{\text{3g}}$ (see Fig. \ref{fig:volume_proof}),
    the network is {\em opt-unsaturated}.

    We now show that $\frac{\partial (\hat{p}(\kappa_{\text{avg}}) 
    \kappa_{\text{avg}})}{\partial \kappa_{\text{avg}}} > 0$
    so that the optimal per-traffic payment is reduced as $\kappa_{\text{avg}}$ decreases.
    We use $\hat{p}(\kappa_{\text{avg}})$ to emphasize the impact of $\kappa_{\text{avg}}.$
    From \eqref{eq:objective_volume} and the condition 
    $R'(\hat{p}(\kappa_{\text{avg}})) = 0,$
  \ifthenelse{\boolean{singlever}}{
    \begin{eqnarray*}
      \label{eq:optimal_price_condition_volume}
      \hat{p}(q)-\eta = - \frac{B(\hat{p})}{B'(\hat{p})}
      = \frac{\hat{p}(1-\theta) \left( 1 - \frac{h(q)}{2-\sigma} \right) }
      { 1 - h(q) },
    \end{eqnarray*}
  }
  {
    \begin{eqnarray*}
      \label{eq:optimal_price_condition_volume}
      \hat{p}(\kappa_{\text{avg}})-\eta & = & - \frac{B(\hat{p})}{B'(\hat{p})} \cr
      & = & \frac{\hat{p}(1-\theta) \left( 1 - \frac{h(\kappa_{\text{avg}})}{2-\sigma} \right) }
      { 1 - h(\kappa_{\text{avg}}) },
    \end{eqnarray*}
  }
    where $h(\kappa_{\text{avg}}) = \frac{\left( \frac{\theta}{\hat{p}(\kappa_{\text{avg}})
    \kappa_{\text{avg}}} \right)^{\frac{1-\sigma}{1-\theta}} }
    { \bar{\Phi}_{\max}^{1-\sigma} } > 0.$
    Differentiating by $\hat{p}$ and rearranging it,
    the derivative $\frac{\partial \hat{p}} { \partial \kappa_{\text{avg}} }$ is as follows:
    \begin{eqnarray*}
      \frac{\partial \hat{p}} { \partial \kappa_{\text{avg}} }
      = - \frac{\hat{p}}{\kappa_{\text{avg}}}
      \frac{ {(1-\sigma)^2} h(\kappa_{\text{avg}})}
      { L(\kappa_{\text{avg}}) + (1-\sigma)^2 h(\kappa_{\text{avg}})},
    \end{eqnarray*}
    where $L(\kappa_{\text{avg}}) = \big(1-h(\kappa_{\text{avg}})\big) \big( \theta (2-\sigma)  - h(\kappa_{\text{avg}}) (1+\theta-\sigma) \big).$
    Thus, the derivative of payment per unit traffic is,
  \ifthenelse{\boolean{singlever}}{
    \begin{eqnarray*}
      \label{eq:optimal_price_first_volume}
      \frac{\partial (\hat{p}(\kappa_{\text{avg}}) \kappa_{\text{avg}})} 
      { \partial \kappa_{\text{avg}} }
      =  \hat{p} + \kappa_{\text{avg}} \frac{\partial \hat{p}} 
      { \partial \kappa_{\text{avg}} }
      = {\hat{p}} \left( \frac{ L(\kappa_{\text{avg}}) }
        { L(\kappa_{\text{avg}}) + (1-\sigma)^2 h(\kappa_{\text{avg}})} \right).
    \end{eqnarray*}
  }
  {
    \begin{eqnarray*}
      \label{eq:optimal_price_first_volume}
      \frac{\partial (\hat{p}(\kappa_{\text{avg}}) \kappa_{\text{avg}})} { \partial \kappa_{\text{avg}} }
      & = &  \hat{p} + \kappa_{\text{avg}} \frac{\partial \hat{p}} { \partial \kappa_{\text{avg}} } \\
      & = & {\hat{p}} \left( \frac{ L(\kappa_{\text{avg}}) }
        { L(\kappa_{\text{avg}}) + (1-\sigma)^2 h(\kappa_{\text{avg}})} \right).
    \end{eqnarray*}
  }
    We want to show that $\frac{\partial (\hat{p}(\kappa_{\text{avg}}) 
    \kappa_{\text{avg}})} { \partial \kappa_{\text{avg}} } > 0.$
    From the condition $R'(\hat{p}(\kappa_{\text{avg}})) = 0,$ we have
    \begin{eqnarray*}
      \frac{1-\frac{h(\kappa_{\text{avg}})}{2-\sigma}}{1-h(\kappa_{\text{avg}})} 
      = \frac{p-\eta}{p(1-\theta)} \leq \frac{1}{1-\theta},
    \end{eqnarray*}
    since $\hat{p} \geq \eta, \eta \geq 0$ and $\theta \in (0,1).$
    Thus, $h(\kappa_{\text{avg}}) \leq \frac{\theta(2-\sigma)}{3-\sigma-\theta}.$
    Applying this condition to $L(\kappa_{\text{avg}}),$
    we have $L(\kappa_{\text{avg}}) > 0$ and therefore,
    \begin{eqnarray*}
     \frac{\partial \hat{p}(\kappa_{\text{avg}})  } { \partial \kappa_{\text{avg}} } < 0 
     ~~~\mbox{ and }~~~
      \frac{\partial (\hat{p}(\kappa_{\text{avg}}) \kappa_{\text{avg}}) } 
      { \partial \kappa_{\text{avg}} } > 0.
    \end{eqnarray*}
    Note that actual payment per unit traffic $\hat{p}(\kappa_{\text{avg}}) \kappa_{\text{avg}}$
    is increasing as $\kappa_{\text{avg}}$ decreases.
    We have proven (iii) holds. \QED


\subsection{Proof of Theorem \ref{thm:volume}}

\noindent \underline{\em (i) Opt-saturated.}
    Recall that 
    the net-utility of a subscriber with ${\Phi}$ is
    \begin{eqnarray*}
\sum_{t \in T} w (t)^{1-\theta} x_{\Phi}^\star (t)^{\theta} -
  p \kappa_{\text{avg}} \sum_{t \in T}  {x}_{\Phi}^\star(t)    
    \end{eqnarray*}
    where
    \begin{eqnarray*}
      {x}_{{\Phi}}^\star (t) = \min \left\{ {\phi} (t) , w(t) \left(\frac{\theta}
      {p \kappa_{\text{avg}}}\right)^{\frac{1}{1-\theta}}  \right\},
    \end{eqnarray*}
    and ${\Phi} = \sum_{t \in T} \phi(t).$
    Therefore the net utility of a user is given by
    \begin{eqnarray*}
      {U}_{\Phi}(x_{{\Phi}}^\star) =
      \left\{ \begin{array}{ll}
          {\Phi}^{\theta} - p \kappa_{\text{avg}} {\Phi}
          & \mbox{if $\Phi < \left( \frac{\theta}{p \kappa_{\text{avg}}} 
          \right) ^{\frac{1}{1-\theta}} $} \\
          \left( 1 - \theta^{\frac{1}{1-\theta}} \right) 
          \left( \frac{\theta}{p \kappa_{\text{avg}}} \right)^{\frac{\theta}{1-\theta}}
          & \mbox{if $\Phi > \left( \frac{\theta}{p \kappa_{\text{avg}}} 
          \right) ^{\frac{1}{1-\theta}}$. }
        \end{array} \right.
    \end{eqnarray*}
    It is clear that the net-utility of a user increases as $p$   decreases.

    Consider the revenue function at the optimal price $p^{\star}$; $R(p^\star)$.
    When a network is opt-saturated, by Proposition~\ref{thm:unimodal_volume}(ii),
    $p^\star = p_0$ such that $A(p_0) = C_{\text{3g}}.$
    From (\ref{eqn:k-alpha-vol-optsat}),  $p^\star$ is dependent of $\kappa_{\text{peak}}$.
    To show that  $R(p^\star)$ increases as $\kappa_{\text{peak}}$ decreases,
    we will show that  $ \frac{\partial R(p^\star)}{\partial \kappa_{\text{peak}}} <0$.
    The first derivative of $R(p^{\star})$ with respect to  $\kappa_{\text{peak}}$ is,
    \begin{eqnarray*}
      \frac{\partial R(p^\star)}{\partial \kappa_{\text{peak}}}
      =  \frac{\partial R(p^\star)}{\partial p^\star}  
      \frac{\partial p^\star}{\partial \kappa_{\text{peak}}}.
    \end{eqnarray*}
    From \eqref{eq:dpdk-volume}, $ p^{\star} $ is an increasing function 
    of $\kappa_{\text{peak}}.$
    Therefore,  it is enough to show
    that $ \frac{\partial R(p^\star)}{\partial p^\star} < 0$.
    We have already  proven that
    $ \frac{\partial R(p^\star)}{\partial p^\star} < 0$ over $\set{P}$ 
    for an opt-saturated network in the proof of Proposition~\ref{thm:unimodal_volume}(ii).




\smallskip
\noindent \underline{\em (ii) Opt-unsaturated.}
    By the same argument of Theorem~\ref{thm:volume}(i),
    net utility of a user and user surplus are increasing as $\kappa_{\text{avg}}$ decreases.

    We now show that revenue is increasing as $\kappa_{\text{avg}}$ decreases.
    We will use  $R(p,\kappa_{\text{avg}}), ~Y(t;p,\kappa_{\text{avg}})$  and
    $\set{P}(\kappa_{\text{avg}})$  to emphasize the impact of $\kappa_{\text{avg}}$.
    Suppose that $\kappa_{\text{avg}}$ is decreased by a factor of $\rho < 1.$
    i.e., $\kappa_{\text{avg}}^{\text{new}} = \rho \kappa_{\text{avg}}.$
    Let  $p^{\text{new}}:= \frac{p}{\rho}$.
    If we show that
    \begin{eqnarray*}
      \forall ~ p \in \set{P}(\kappa_{\text{avg}}) 
      ~~ \Rightarrow ~~ p^{\text{new}} \in \set{P}\left({\kappa_{\text{avg}}^{\text{new}}}\right)
    \end{eqnarray*}
    and $R\left(p^{\text{new}}, \kappa_{\text{avg}}^{\text{new}} 
    \right)> R(p,\kappa_{\text{avg}})$ hold,
    then
    \begin{eqnarray*}
      R(p^{\star},\kappa_{\text{avg}})  
      <  R\left(\frac{p^{\star}}{\rho}, \kappa_{\text{avg}}^{\text{new}} \right) \leq
      \max_{ p \in \set{P}( {\kappa_{\text{avg}}^{\text{new}}}) }  
      R(p, \kappa_{\text{avg}}^{\text{new}})
    \end{eqnarray*}
    where $p^{\star}$ is the optimal price with $\kappa_{\text{avg}}^{\text{new}}$
    which implies that the maximum revenue is increasing as $\kappa_{\text{avg}}$ is decreasing.

    We show that  $ p^{\text{new}} \in \set{P} 
    \big( \kappa_{\text{avg}}^{\text{new}}  \big)   $;
    $R(p^{\text{new}},\kappa_{\text{avg}}^{\text{new}}) > 0$ and
    $\max_{t \in {T}} Y(t;p^{\text{new}},\kappa_{\text{avg}}^{\text{new}}) \leq C_{\text{3g}}.$
    By 
    \eqref{eq:net_utility_volume}, it is obvious that
    \begin{eqnarray*}
      {U}_{p, \kappa_{\text{avg}}} (x)  
      =   {U}_{p^{\text{new}}, \kappa_{\text{avg}}^{\text{new}}} (x).
    \end{eqnarray*}
    Since the user optimization problems are identical,
    the total traffics are the same, ${\bm{x}^{\star}}^{\text{new}} = \bm{x}^{\star}$.
    Then the corresponding 3G traffic $\bm{y}^{\text{new}}$ and $\bm{y}$ satisfy
    \begin{eqnarray*}
      \sum_{t \in {T}} {y}^{\text{new}} = \kappa_{\text{avg}}^{\text{new}} 
      \sum_{t \in {T}} {x}^{\star}(t)  =
      \rho \sum_{t \in {T}} y(t) < \sum_{t \in {T}} y(t).
    \end{eqnarray*}
    Therefore
  \ifthenelse{\boolean{singlever}}{
    \begin{eqnarray}
      \sum_{t \in {T}} Y \left(t;p^{\text{new}}, \kappa_{\text{avg}}^{\text{new}} \right)
      = \rho \sum_{t \in {T}} Y(t;p, \kappa_{\text{avg}})  \label{eqn:Y-new}
      < \sum_{t \in {T}} Y(t;p, \kappa_{\text{avg}}),   \nonumber\
    \end{eqnarray}
  }
  {
    \begin{eqnarray}
      \sum_{t \in {T}} Y \left(t;p^{\text{new}}, \kappa_{\text{avg}}^{\text{new}} \right)
      & = & \rho \sum_{t \in {T}} Y(t;p, \kappa_{\text{avg}})  \label{eqn:Y-new} \\
      & < & \sum_{t \in {T}} Y(t;p, \kappa_{\text{avg}}),   \nonumber\
    \end{eqnarray}
  }

    Consider the corresponding revenue function $R(p, \kappa_{\text{avg}})$ and $R(p^{\text{new}},\kappa_{\text{avg}}^{\text{new}})$.
  \ifthenelse{\boolean{singlever}}{
    \begin{eqnarray*}
      R\left(p^{\text{new}},q^{\text{new}}\right)
      & = &
      \left(\frac{p}{\rho} -\eta\right)
      \sum_{t \in {T}}  Y \left(t; p^{\text{new}},q^{\text{new}} \right)
      = (p  -\rho \eta) \sum_{t \in {T}} Y \left(t; p,q \right)   ~(\because (\ref{eqn:Y-new})) \\
      & > &   R(p,q) ~(\because p \in \set{P}({q})),
    \end{eqnarray*}
  }
  {
    \begin{eqnarray*}
      R\left(p^{\text{new}},\kappa_{\text{avg}}^{\text{new}}\right) 
      & = &
      \left(\frac{p}{\rho} -\eta\right)
      \sum_{t \in {T}}  Y \left(t; p^{\text{new}},\kappa_{\text{avg}}^{\text{new}} \right) \\
      & = & (p  -\rho \eta) \sum_{t \in {T}} Y \left(t; p,\kappa_{\text{avg}} \right)   ~(\because (\ref{eqn:Y-new}))\\
      & > &   R(p,\kappa_{\text{avg}}) ~(\because p \in \set{P}({\kappa_{\text{avg}}})),
    \end{eqnarray*}
  }
    where $R(p,\kappa_{\text{avg}}) = (p  - \eta) \sum_{t \in {T}} Y \left(t; p,\kappa_{\text{avg}} \right).$
    Now we have shown that 
    $R\left(p^{\text{new}},\kappa_{\text{avg}}^{\text{new}}\right)  > 
    R(p,\kappa_{\text{avg}}) > 0.  $
    Hence $\rho^{-1} p \in \set{P}\left({\kappa_{\text{avg}}^{\text{new}}}\right).$ \QED





\end{document}